\documentclass[manuscript]{acmart}

\usepackage{url}
\usepackage{enumitem}

\usepackage{amsmath,amssymb,amsfonts}
\usepackage{algorithmic}
\usepackage{graphicx}
\usepackage{textcomp}
\usepackage[table,xcdraw]{xcolor}
\usepackage{booktabs}

\usepackage{multirow}
\usepackage[table,xcdraw]{xcolor}
\usepackage[normalem]{ulem}

\usepackage[most]{tcolorbox}
\newcounter{findingcounter}
\renewcommand{\thefindingcounter}{\Roman{findingcounter}} 
\newcommand{\rqone}{What feedback did developers provide on CodeRabbit's code reviews?}
\newcommand{\rqtwo}{How do different types of concerns raised in CodeRabbit's code reviews relate to developer feedback?}
\newcommand{\rqthree}{Can we predict developer's feedback to CodeRabbit's code reviews?}

\usepackage{listings}
\usepackage{color}

\definecolor{myred}{HTML}{FE7777}

\definecolor{codegray}{rgb}{0.5,0.5,0.5}
\definecolor{backcolour}{rgb}{0.95,0.95,0.92}
\definecolor{lightgray}{HTML}{EAEAEA}

\newtcolorbox{summarybox}{
    colback=gray!10,
    colframe=gray!30,
    sharp corners,
    boxrule=0.5pt,
    left=5pt, right=5pt,
    top=5pt, bottom=5pt,
    before skip=10pt,
    after skip=10pt
}

\newcommand{\summary}[1]{
    \stepcounter{findingcounter}
    \begin{summarybox}
        \textbf{Key Finding \thefindingcounter:} #1
    \end{summarybox}
}

\title{Is Agentic Code Review Helpful? Mining Developers' Feedback to CodeRabbit Reviews in the Wild}

\author{Hong Yi Lin}
\email{holin2@unimelb.edu.au}
\orcid{0009-0004-5368-8897}
\affiliation{%
  \institution{The University of Melbourne}
  \city{Melbourne}
  \state{Victoria}
  \country{Australia}
}

\author{Mingzhao Liang}
\email{mingchiu.leung@gmail.com}
\orcid{0009-0009-7315-9497}
\affiliation{%
  \institution{The University of Melbourne}
  \city{Melbourne}
  \state{Victoria}
  \country{Australia}
}

\author{Patanamon Thongtanunam}
\email{patanamon.t@unimelb.edu.au}
\orcid{0000-0001-6328-8839}
\affiliation{%
  \institution{The University of Melbourne}
  \city{Melbourne}
  \state{Victoria}
  \country{Australia}
}

\author{Kla Tantithamthavorn}
\orcid{0000-0002-5516-9984}
\email{chakkrit@monash.edu}
\affiliation{%
  \institution{Monash University}
  \city{Melbourne}
  \state{Victoria}
  \country{Australia}
}

\renewcommand{\shortauthors}{Lin et al.}

\begin{document}



\begin{abstract}
Agentic code review, where autonomous agents provide code review comments on pull requests, is increasingly integrated into development workflows, yet there is limited empirical evidence on how developers respond to such comments in practice. 
In this paper, we present an empirical study of agentic code reviews using CodeRabbit as a case study. 
Through an empirical study of 31,073 pairs of code reviews and developer feedback from 10,191 pull requests across 239 GitHub repositories,
our results show that agentic reviews receive mixed reception: 36.4\% were accepted and 7.3\% triggered discussion, while 56.3\% were rejected. 
Rejections were primarily associated with invalid suggestions that were false positives, redundant, or out of scope, as well as misalignment with developer intent and coding practices.
We further found that agentic reviews tend to focus more on functional concerns than evolvability-related comments, yet they were more likely to be invalid.
To improve effectiveness in review practices, we explored various LLM-based approaches for predicting review rejection. 
We found that lightweight learning-based methods achieve up to 76\% F1 score, suggesting learnable patterns exist between code reviews and their corresponding feedback.
Our results highlight the current state of CodeRabbit's agentic code reviews, showing opportunity gaps for improvement, as well as shortcomings hindering its effectiveness.
\end{abstract}

\maketitle

\section{Introduction}
\begin{figure}[t]
  \centering
  \includegraphics[width=0.65\columnwidth]{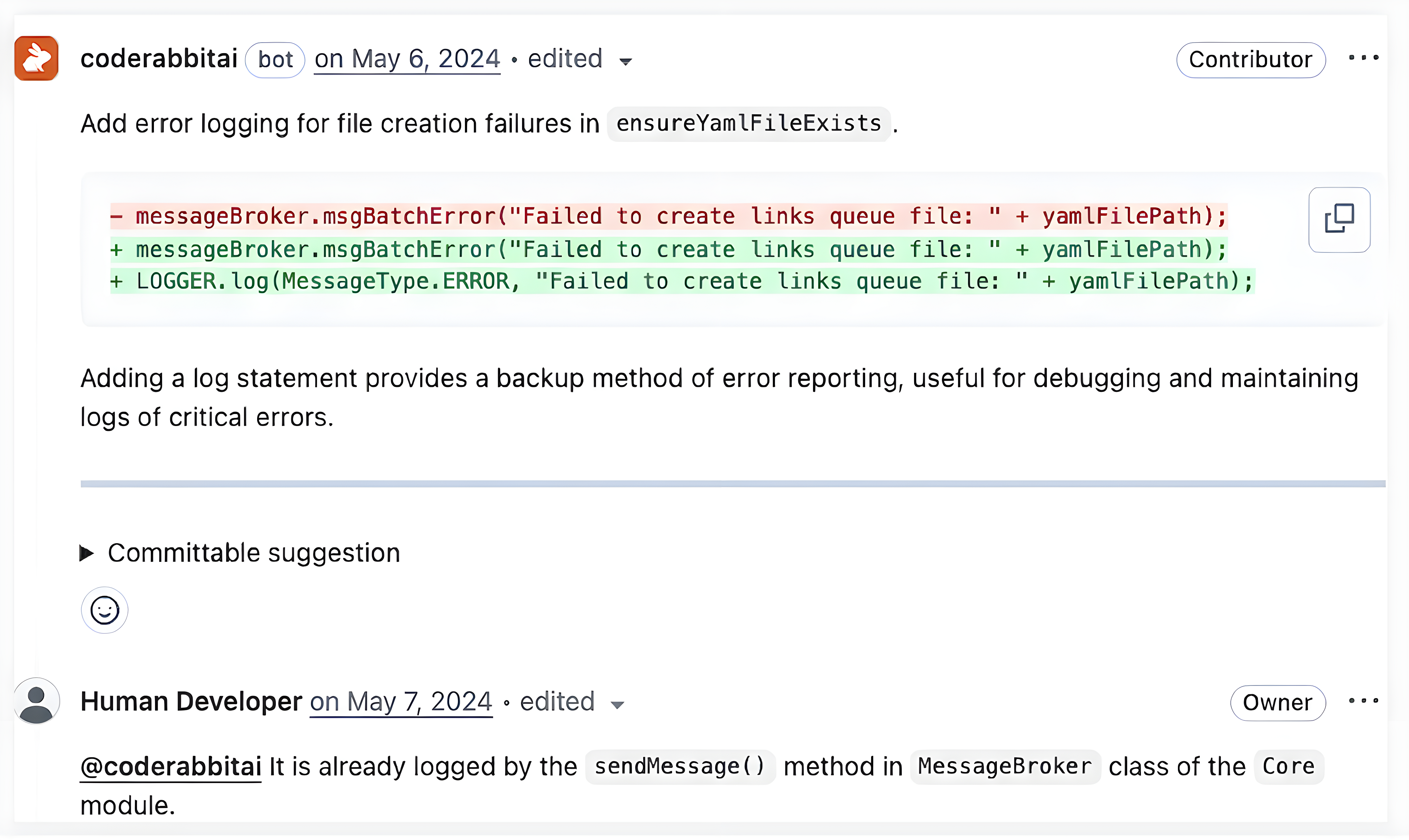}
  \caption{An Example of a Developer Responding to an Agentic Code Review Provided by CodeRabbit}
  \label{fig:example_1}
  \Description{coderabbit example}
\end{figure}

The modern code review is an essential software quality assurance practice that helps ensure the maintainability and overall quality of a software system.
Recently, the field has witnessed a transformative shift from human review workflows to hybrid human–AI collaborations in the code review practices \cite{tantithamthavorn2026rovodev,coderabbit2026}.
In this new paradigm, Large Language Models (LLMs) are no longer just passive completion engines, they function as autonomous reviewer agents capable of providing real-time, inline feedback on pull requests (PRs).
\textit{CodeRabbit}~\cite{coderabbit2026,coderabbit2026pipeline} is one of the most popular agentic code review tools, aiming to provide context-aware, inline feedback directly within a PR (see Figure \ref{fig:example_1}).
Unlike conventional ``static-call'' LLM-based approaches which often analyse code snippets in isolation \cite{li2022automating, 10378848, 10.1145/3762183}, agentic tools operate with higher levels of agency. 
These agents are designed to traverse project-level context, interpret repository-wide dependencies, and orchestrate external static analysis tools (e.g., linters or security scanners) to synthesise multi-faceted review comments.
While these tools promise to reduce reviewer fatigue, their rapid adoption has outpaced empirical understanding of their actual utility in production environments.

While research has a solid understanding of the review quality and challenges in the human code reviews~\cite{10.1145/2884781.2884840,9425884,10.1145/3611643.3616245}, the recent shift toward agentic code review tools has created a knowledge gap in this emerging AI‑integrated code review practice.
Although recent studies based on benchmarking datasets have pointed out that automated code reviews can also be noisy~\cite{11025607} or even hallucinated~\cite{liu-etal-2025-hallucinations}, i.e.,  generating reviews that are not aligned with the code changes or their context~\cite{tantithamthavorn2026hallujudge}, 
they focused on the review as a static artifact, without considering the subsequent developer feedback. 
A developer's explicit feedback is a vital indicator of review effectiveness in practice~\cite{tantithamthavorn2026hallujudge,tantithamthavorn2026rovodev,VijayvergiyaSBZ24}. 
For example, Figure~\ref{fig:example_1} shows that the human developer responded to CodeRabbit's suggestion by noting it had already been implemented, indicating that CodeRabbit's comment may not be useful in this case.
An agentic code review would be considered helpful if it provides valid suggestions or raises concerns that lead to code improvements or stimulate developer collaboration~\cite{7180075, hasan2021using}.
Unfortunately, developer feedback to agentic code reviews remain largely unexplored, raising questions about whether these agentic code review tools are helpful to software developers in real-world use cases.

In this paper, we conducted an empirical study to examine the helpfulness of CodeRabbit's \textbf{\textit{agentic code reviews}} in practice.
To do so, we manually analysed its agentic code reviews along with their corresponding developer feedback to understand how developers responded and engaged with these reviews.
We also analysed the association between developer feedback and the concern types raised in agentic code reviews.
Finally, to reduce the amount of non-helpful code reviews in the workflow, we explored LLM‑based approaches for early predicting developer rejections of CodeRabbit's agentic code reviews.
Specifically, we investigated three LLM-based approaches, i.e., direct prompting with large LLMs, low-rank adaption with small-scale LLMs, and fine‑tuning  encoder-only models from five state‑of‑the‑art model families (GPT, Llama, Qwen, DeepSeek, and ModernBERT). 
Based on 31,073 pairs of CodeRabbit's code review comments and developer feedback from 10,191 pull requests across 239 GitHub repositories spanning across ten of the most popular programming languages, we answer the three following research questions:

\begin{enumerate}[label=\textbf{(RQ\arabic*}),leftmargin=*]

\item \textbf{\rqone} \\
We found that 36.4\% of agentic code reviews were accepted by developers, either leading to code changes or resulting in a new issue ticket. 
Another 7.3\% of the reviews initiated a discussion among developers. 
Nonetheless, 56.3\% of the code reviews were still rejected by developers.
The common reasons for rejections were misalignment with coding practices and suggestions being invalid.
These results highlight that a substantial amount of code reviews that did not assist developers in improving the code changes.
On a postive note, the proportion of accepted reviews were increasing overtime (12.5\% since release), indicating an improvement in the tool itself.

\item \textbf{\rqtwo} \\ 
While CodeRabbit's agentic code reviews raised concerns in both functionality and evolvability, they were often rejected with a rate of 56.2\% and 56.7\%, respectively. 
Specifically, CodeRabbit has increasingly been attempting to identify more functional defects, accounting for 43.3\% of all code reviews.
However, 35.8\% of them were rejected due to being invalid. 
In contrast, evolvability‑related reviews were mostly rejected due to misalignment in coding practices (31.6\%), rather than being invalid (25.1\%).
These results highlight flaws in CodeRabbit's technical proficiency when detecting functional defects and its lack of alignment with developer practices.

\item \textbf{\rqthree} \\
When early predicting whether CodeRabbit's code review will be rejected, direct prompting with large LLMs achieved an F1 score of up to 64\%.
In contrast, learning‑based approaches, i.e., low‑rank adaptation and fine‑tuning, achieved F1 scores of up to 76\% and 68\%, respectively.
These results suggest that small-scale LLMs can learn meaningful patterns between the code reviews and their associated feedback, allowing them to act as an effective quality gate that reduces the amount of low quality reviews that reach the developer.
\end{enumerate}

\textbf{Implications.} Despite the promising capabilities and widespread adoption of the CodeRabbit agentic code review tool, our findings reveal notable limitations in practice. 
Such non‑helpful reviews that were eventually rejected could introduce additional overhead, as developers spend extra effort to validate them. 
Nevertheless, we identified potential lightweight techniques  for early prediction of non‑helpful code reviews that are likely to be rejected, which can be used as an effective quality gate. 
Integrating such predictive approaches into the code review workflow could reduce non-helpful reviews, ultimately improving perceived reliability and fostering long‑term trust in such tools.

\textbf{Contributions \& Novelty.}  To the best of our knowledge, this paper makes the following contributions:

\begin{itemize}
\item We are the first to provide a large collection of 31,073 pairs of CodeRabbit's code review comments and developer feedback from 10,191 pull requests across 239 GitHub repositories spanning across ten programming languages.

\item We present a large-scale and systematic analysis of the distribution of developer feedback to CodeRabbit's agentic code reviews, the underlying reasons that lead to their rejection, how these feedback patterns vary based on the different types of concerns raised in the review, and how they have evolved over time.

\item We demonstrate that small-scale LLMs trained on developer feedback patterns can be used to early predict whether an agentic code review is likely to be rejected, offering a practical lightweight quality gate.
\end{itemize}

\begin{figure}[!htbp]
    \centering
    \includegraphics[width=0.6\linewidth]{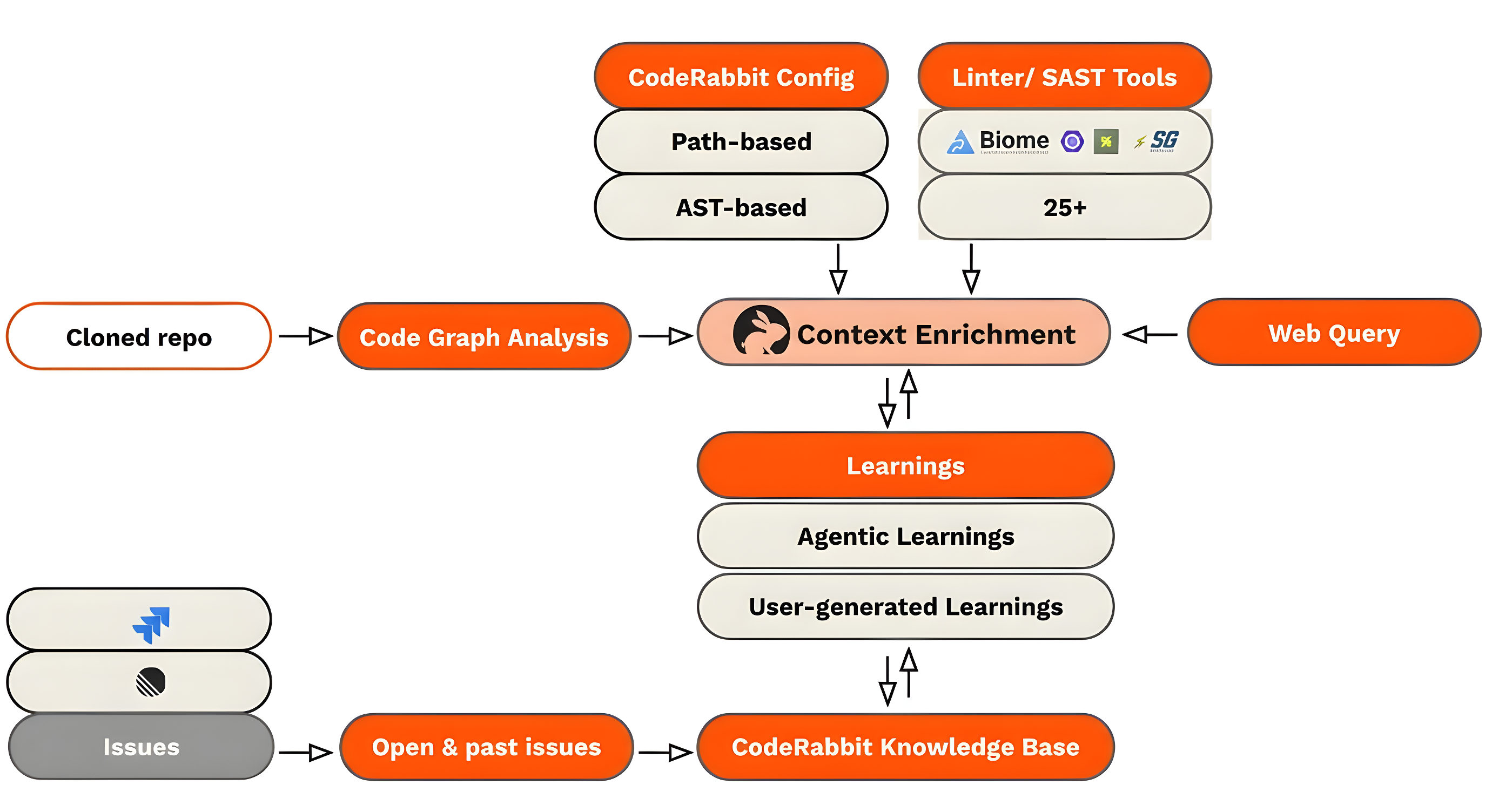}
    \caption{An Overview of CodeRabbit's Agent Architecture}
    \label{fig:coderabbit}
    \Description{architecture}
\end{figure}

\section{Background and Related Work}

In this section, we briefly describe CodeRabbit and the literature related to both human and automated code reviews.

\subsection{CodeRabbit's Agentic Code Review}
CodeRabbit~\cite{coderabbit2026,coderabbit2026pipeline} employs an agentic architecture designed to move beyond traditional linear static LLM pipelines by emphasising deep contextual reasoning and multi-source data integration. 
As of March 2026, CodeRabbit is the most installed app in the GitHub market under the code review category~\cite{githubmarketplacecodereview} with more than 200K installations~\cite{githubcoderabbit}.

As illustrated in the provided architectural workflow (see Figure \ref{fig:coderabbit}), the core of the system is the context enrichment engine, which synthesises disparate inputs to provide the CodeRabbit agent with a comprehensive understanding of the codebase. 
This enrichment process begins with a code graph analysis of the cloned repository, enabling the system to understand structural dependencies and control flows. 
Simultaneously, the engine ingests configuration data, both path-based and AST-based (Abstract Syntax Tree), alongside diagnostic outputs from a suite of over 25 linters and static analysis security testing (SAST) tools, such as Biome and Semgrep.
To ensure the review is grounded in the specific evolution of the project, the architecture incorporates a bi-directional feedback loop with the CodeRabbit knowledge base. 
This repository of intelligence is fed by open \& past issues from platforms like Jira and GitHub, as well as a ``Learnings'' layer that distinguishes between agentic learnings (autonomous refinements by the AI) and user-generated learnings (explicit feedback from human developers). 
Furthermore, CodeRabbit is capable of performing web queries to resolve external documentation or library dependencies in real-time. 
The orchestration of these components is designed to transition from simple syntax checking to a sophisticated reasoning framework targeting complex logic flaws and architectural inconsistencies that traditional static analysis might not be able to capture.

\subsection{Code Review Quality \& Challenges}

\textbf{Human Code Review.}
The effectiveness of human-conducted code reviews has been extensively studied. 
A survey of the open-source Mozilla Core project found that developers considered reviews based on limited code knowledge to be superficial and not useful~\cite{10.1145/2884781.2884840}. 
Microsoft developers reported reviews that merely asked questions, praised code, suggested out of scope work, or flagged unrelated issues were unhelpful~\cite{7180075}.
Other studies at the same institution noted low quality reviews that only focused on obvious errors~\cite{6606617}. 
Research on OpenStack further identified anti-patterns such as confused, conflicting, brief, shallow, and toxic reviews~\cite{9425884}, whilst reviews in the OpenDev Nova project were found to raise invalid concerns~\cite{turzo2024makes}. 
Although human-conducted code reviews have been well-studied in practice, developer feedback on agentic code reviews remain largely unexplored in real-world settings.

Several studies have investigated methods for automatically identifying low-quality human code reviews. 
These approaches include traditional machine learning classifiers that leverage a combination of socio-technical and keyword-based features~\cite{7180075,hasan2021using,rahman2017predicting}, as well as encoder-only models that operate directly on code review comments~\cite{10.1145/3611643.3616245}.
In terms of their shortcomings, these methods either depend on post-review features, limiting their applicability for early prediction~\cite{7180075,hasan2021using}; rely on human-specific attributes that are unavailable for automated agents (e.g., reviewer experience)~\cite{rahman2017predicting}; or focus primarily on surface-level linguistic signals as proxies for review quality~\cite{10.1145/3611643.3616245}.

\textbf{Automated Code Review.}
To alleviate the workload of human reviewers~\cite{10.1145/2950290.2950323,7202946}, the field of automated code review~\cite{10378848} seeks to develop language models capable of generating high quality code review comments. 
Early approaches often rely on techniques such as task-specific pretraining~\cite{LiYJYLHLZ22,10.1145/3510003.3510621,li2022automating} or specialised fine-tuning~\cite{10.1145/3762183,10.1145/3643991.3644910} to enhance model performance.
However, these earlier models were primarily evaluated on offline test sets, where effectiveness was measured through text-based similarity to reference comments and manual assessment of quality attributes. 
Due to their limited capabilities and reliance on a “static-call” setup i.e., only inspecting a code hunk, such models were either never deployed in real-world settings, or restricted to addressing only superficial coding practices~\cite{VijayvergiyaSBZ24}.
In contrast, modern LLM-based tools, while still largely static, have been integrated into certain developer workflows, shifting the focus of evaluation to their practical effectiveness in eliciting subsequent code changes~\cite{11495549}. 
However, they do not reflect the state-of-the-art agentic paradigm that practictioners are shifting towards~\cite{tantithamthavorn2026rovodev, GoldmanLPTTWZBJSJBJW25}.

As we transition to the agentic paradigm, increasingly capable code review systems have emerged~\cite{tantithamthavorn2026hallujudge, tantithamthavorn2026rovodev, GoldmanLPTTWZBJSJBJW25}, leveraging project- and repository-level context, as well as external static analysis tools. 
Prior research from industry have examined aspects such as resolution rates using change-proximity-based proxies~\cite{GoldmanLPTTWZBJSJBJW25}, perceptions of in-house developers through small-scale surveys~\cite{11121707}, detection of hallucinated reviews~\cite{tantithamthavorn2026hallujudge}, and broader impacts on the code review process, such as pull request cycle time~\cite{tantithamthavorn2026rovodev}.
However, there lacks a large-scale analysis into how developers in the open source actually respond to the various types of agentic code reviews at the instance-level, as well as the underlying reasons behind their feedback.
Motivated by evidence that code reviews from LLM-powered tools can be of low quality~\cite{liu-etal-2025-hallucinations, tantithamthavorn2026hallujudge}, we assess the helpfulness of agentic code reviews based on large-scale developer feedback in practice, identify the underlying reasons, and conduct a systematic evaluation of LLM-based approaches for early feedback prediction, which can be utilised for preventing low quality comments from reaching the developer.

\section{Study Design}
In this section, we present the motivations of our research questions, and the data preparation process for our study.

\subsection{Research Questions}
\begin{enumerate}[label=\textbf{(RQ\arabic*}),leftmargin=*]

\item \textbf{\rqone} \\
\underline{Motivation:} While prior work has examined the quality of human‑written reviews and how developers react to them in reality~\cite{10.1145/2884781.2884840,turzo2024makes,6606617,9425884,turzo2024makes}, the assessment of AI code reviews has largely been limited to offline evaluations~\cite{10378848,10.1145/3762183,LiYJYLHLZ22,li2022automating,10.1145/3510003.3510621, tantithamthavorn2026hallujudge}. Thus, it remains unclear to what extent agentic reviews are helpful in practice and why developers may reject them. Therefore, we investigate actual developers' feedback on CodeRabbit's code reviews to determine their practical usefulness and the concerns that lead developers to dismiss them.

\item \textbf{\rqtwo} \\
\underline{Motivation:} Several studies have investigated the capabilities of both human and automated code review by examining the distribution of comment types~\cite{GoldmanLPTTWZBJSJBJW25}. 
However, it remains unclear which types of reviews generated by CodeRabbit receive positive feedback and which types are more likely to be rejected by developers. 
Therefore, we set out to investigate the types of reviews provided by the tool and how the distribution of feedback varies.

\item \textbf{\rqthree} \\
\underline{Motivation:} Given that the utility of agentic code reviews profoundly influences developer experience, implementing a quality gate is essential for mitigating non-helpful code reviews. 
However, it remains unclear which techniques are most effective for predicting whether a code review would ultimately be rejected. 
We therefore systematically evaluate a range of state-of-the-art techniques for early rejection prediction.

\end{enumerate}

\subsection{Data Collection and Preparation}

\begin{table*}[!htbp]
\center
  \caption{Summary Statistics of Our Dataset}
  \label{tab:data_summary}
  \resizebox{0.55\textwidth}{!}{%
    \Large
    \sffamily
    \rowcolors{2}{white}{gray!15}
    \begin{tabular}{|l|r|rrrr|}
      \hline
      \multicolumn{1}{|l|}{\cellcolor[HTML]{EFEFEF}\textbf{Programming}}
      & \cellcolor[HTML]{EFEFEF}\textbf{\#Projects}
      & \multicolumn{4}{c|}{\cellcolor[HTML]{EFEFEF}\textbf{Agentic Reviews}} \\
      \multicolumn{1}{|l|}{\cellcolor[HTML]{EFEFEF}\textbf{Language}}
      & \cellcolor[HTML]{EFEFEF}\textbf{}
      & \cellcolor[HTML]{EFEFEF}\textbf{\#}
      & \cellcolor[HTML]{EFEFEF}\textbf{Min. Len}
      & \cellcolor[HTML]{EFEFEF}\textbf{Med. Len}
      & \cellcolor[HTML]{EFEFEF}\textbf{Max. Len} \\
      \hline
      Go &  48 & 6,393 & 20 & 170 & 2,483 \\
      TypeScript & 45 & 6,751 & 17 & 184 & 1,921 \\
      Python &  34 & 3,691 & 1 & 202 & 1,289 \\
      JavaScript &  27 & 2,768 & 16 & 177 & 1,280\\
      Rust &  25 & 2,081 & 10 & 203 & 2,911 \\
      C\# &  24 & 2,734 & 21 & 174 & 1,133 \\
      Java & 22 & 5,836 & 11 & 165 & 3,437 \\
      Kotlin &  7 & 383 & 33 & 196 & 1,188 \\
      PHP &  5 & 256 & 23 & 167 & 900 \\
      HTML &  2 & 180 & 48 & 176 & 953 \\
      \hline
      \rowcolor{gray!15}
      \textbf{Overall}
      & \textbf{239}
      & \textbf{31,073}
      & \textbf{1}
      & \textbf{181}
      & \textbf{3,437} \\ \hline
      \rowcolor{white}
      \multicolumn{6}{l}{Len. = Length of a review comment measured by word count.}
    \end{tabular}
  }
\end{table*}

\textbf{Raw Data Collection.}
To identify GitHub projects that use the CodeRabbit agentic code review tool, we queried the GitHub REST API for pull requests (PRs) containing review comments that directly mention CodeRabbit (i.e., @coderabbitai).
These direct mentions provide a clear signal of the interactions between developers and CodeRabbit, allowing us to pinpoint projects that are directly utilising CodeRabbit in their workflows. 
Our data spans from September 19, 2023, CodeRabbit's release date, to September 30, 2025, the date of our data collection.
This process yielded \textbf{119,115} PRs from \textbf{21,959} distinct public GitHub projects, forming our initial pre-filtered corpus.

\textbf{GitHub Project Selection.}
To ensure data quality and compliance with reuse requirements, we applied three selection criteria at the project level.
First, we selected only the projects whose primary language is among the ten most popular programming languages. This helps ensure that our findings generalise across commonly-used programming languages.
Based on our initial corpus, we selected the 16,966 projects written in TypeScript, Python, JavaScript, Java, Go, Rust, PHP, C\#, HTML, and Kotlin.
Collectively, these projects represent 77\% of the projects in our initial dataset.
Second, we selected projects based on their software licenses. This step helps ensure that the associated code changes and PRs are publicly accessible and reusable~\cite{li2022automating}. We retained only projects that declare permissive and widely adopted licenses, i.e., the MIT License and the Apache License 2.0, which explicitly allow analysis and redistribution of derived artifacts. After applying this filter, 6,356 projects remained.
Finally, we excluded projects that may use CodeRabbit only experimentally, i.e., those that interacted with CodeRabbit only a few times and then discontinuing use.
To determine this, we examined the distribution of projects based on the number of PRs that mention \texttt{@coderabbitai}. 
We found that, on average, a project contains around 20 such PRs.
Therefore, we retained only projects with more than 20 PRs. This final selection step resulted in \textbf{239} projects, forming a final set of selected projects.

\textbf{CodeRabbit Review Comment Selection.}
To ensure that we capture agentic reviews and developers' feedback regardless of whether a direct mention was used, we retrieved all PRs and reviews for the final set of selected projects.
This process yielded a total of 781,966 PRs and 1,953,469 reviews. 
To focus specifically on agentic code reviews, we selected only the reviews provided by CodeRabbit that inserted directly to the code changes (see Figure \ref{fig:example_1}). 
We focus on the initial review in the discussion thread to capture the primary review by CodeRabbit~\cite{11495549}.
The tool is generally not utilised in a multi-turn setting, and only 9\% of discussions have more than 2 replies from any GitHub account.
As a result, we retained 332,693 reviews authored by CodeRabbit. 
Since we aim to analyse the developer feedback to agentic reviews, we excluded CodeRabbit reviews that did not receive any developer replies. 
Based on this criterion, we obtained \textbf{31,073} CodeRabbit reviews for our study.
Table \ref{tab:data_summary} shows the summary statistics of the final dataset.

\section{Results}
In this section, we present the analysis approach and results for each of our research questions.

\subsection*{(RQ1) \rqone}
\textbf{Approach.} 
To answer our research question, we first conducted lightweight coding~\cite{10.1145/2568225.2568260} on 297 sampled agentic code reviews, representing 10\% of the dataset through stratified sampling across repositories and programming languages.
For each sample, we examined the code diff under review, the inline agentic code review comment, as well as the subsequent developer feedback. 
Our goal was to understand how agentic code reviews were being received by developers using the tool. 
During this process, we recorded both the developer feedback and the underlying reason.
We performed an open coding exercise to group the feedback and reasons into higher-level themes based on the following four steps.

\begin{enumerate}
    \item The first author synthesised the scenario in each sample into concise summaries.
    \item The first author grouped the samples based on thematic similarities of their summaries.
    \item The second author reviewed the groups and raised any disagreements with the first author. When disagreements could not be resolved by the first two authors, the third author examined the discrepancies and facilitated discussion until a consensus was reached.
    \item The first author refined the groups according to the agreed-upon resolutions.
\end{enumerate}

Steps 2-4 were iterated for two rounds, after which no new groups emerged.
Finally, the second author validated all samples to confirm the correctness of the groupings.
All manual coders had relevant experience, i.e., the first two coders previously worked as professional software engineers, and the third coder has extensive experience in software engineering research.
After establishing the categories, we evaluated GPT-5.1's ability to classify samples according to our manually derived labels. 
We found that it can achieve substantial agreement (Cohen's $\kappa = 0.74$) with manual coders. 
Consequently, we used it to annotate the remaining 30,776 agentic code reviews.
The prompts are shown in Figure~\ref{fig:feedback_categorisation}.
In addition to reporting the distribution over the entire dataset, we also statistically examine monthly distributional trends from CodeRabbit's release on September 19, 2023, to September 30, 2025, the date of our data collection. 
We perform the time series analysis at a monthly granularity to align with CodeRabbit's release frequency, including changes to its underlying frontier models, allowing us to capture distributional shifts associated with these updates over time.
Specifically, we use the Mann-Kendall trend test~\cite{mann1945nonparametric,19521603271} and report Kendall's Tau, the p-value, and Sen's slope~\cite{sen1968estimates}.

\begin{figure}[!htbp]
\begin{tcolorbox}[colback=gray!5!white, colframe=lightgray, coltitle=black, fonttitle=\bfseries\sffamily\footnotesize , fontupper=\sffamily\footnotesize, title=High Level Developer Feedback Labelling Prompt]

You are an expert software engineering researcher analysing human-AI interactions.
Given this agent-generated code review by CodeRabbit AI, label the subsequent feedback from the human developer into exactly one of the three predefined categories.\\

The feedback should be one of the following three types.\\

\textbf{1) Accepted Suggestion:} This is a positive feedback signal, where the developer will find the code review useful and choose to incorporate the suggestion into a code change e.g., submitting a new commit, opening a corresponding PR.\\
\textbf{2) Triggered Discussion:} This is the case where there is no direct action; rather, the code review will trigger further discussion between human code reviewers and developers. Usually, the human who initially read the code review requires further confirmation regarding the validity of the code review from a better-informed team member.\\
\textbf{3) Rejected Suggestion:} This is a negative feedback signal, where the code review will either be rejected due to the developer's own preferences and intentions or due to the code review itself being incorrect e.g., false positive issue, redundant suggestion, or out of scope.\\

If there was no intention of action to incorporate the suggestion or agreement with the suggestion, it CANNOT BE labelled as Accepted Suggestion. 
Another human user MUST BE called by `Human Feedback:' using @ for it to be Triggered Discussion. Feedback that uses @coderabbitai or does not use @ are directed towards the agent and MUST NOT BE labelled as Triggered Discussion. Otherwise, it MUST BE a Rejected Suggestion. The rejection can be either explicitly or implicitly conveyed. 
The developer feedback is denoted by `Human Feedback:', this should be the only target of the labelling. If exists, any subsequent agent reply to the developer feedback is denoted by `coderabbitai:', this is not the feedback itself and should not be the target of the labelling.\\

{[}CODE\_REVIEW{]}\\
\{code\_review\}\\
{[}\textbackslash CODE\_REVIEW{]}\\

{[}FEEDBACK{]}\\
\{human\_reply\}\\
{[}\textbackslash FEEDBACK{]} \\

Directly answer with one of the 1, 2, 3 symbols:
\end{tcolorbox}

\begin{tcolorbox}[colback=gray!5!white, colframe=lightgray, coltitle=black, fonttitle=\bfseries\sffamily\footnotesize , fontupper=\sffamily\footnotesize, title=Developer Rejection Reason Labelling Prompt]

You are an expert software engineering researcher analysing human-AI interactions.
Below is a code review generated by CodeRabbit AI that was ultimately rejected by the human developer. 
Given this code review and corresponding developer feedback, label the primary reason for rejection into exactly one of the five predefined categories.\\

The feedback is denoted by `Human Feedback:', this should be the target of the labelling. 
If exists, any subsequent agent reply to the human feedback is denoted by `coderabbitai:', this is not the feedback itself and should not be the target of the labelling. \\

The rejection reason should be one of the following five types. \\

\textbf{1) Intended Design Trade-off:} Suggesting to improve a suboptimal code implementation that was intentionally chosen by the developer. \\
\textbf{2) Developer Preference:} Suggesting a code improvement that diverges from the developer’s personal coding preferences. \\
\textbf{3) False Positive Issue:} Suggesting to fix an issue with the code that was actually misidentified and did no constitute an actual defect.\\
\textbf{4) Redundant Suggestion:} Suggesting a code improvement or identifying an issue that has already been addressed.\\
\textbf{5) Out of Scope:} Suggesting a code improvement that was beyond the objective of the pull request that was undergoing the code review. \\

{[}CODE\_REVIEW{]}\\
\{code\_review\}\\
{[}\textbackslash CODE\_REVIEW{]}\\

{[}FEEDBACK{]}\\
\{human\_reply\}\\
{[}\textbackslash FEEDBACK{]} \\

Directly answer with one of the 1, 2, 3, 4, 5 symbols:
\end{tcolorbox}

\caption{\normalsize Developer Feedback Labelling Prompts}
\label{fig:feedback_categorisation}
\Description{feedback_categorisation}
\end{figure}

\textbf{Results.} Table~\ref{tab:feedback_categories} shows three high level categories of developer feedback, i.e., accepting the suggestion, triggering a discussion, or rejecting the suggestion.
We find that 36.4\% of agentic code reviews were accepted by developers, whilst 7.3\% of agentic reviews triggered a discussion amongst human developers. 
Nevertheless, 56.3\% of agentic code reviews were still rejected by developers. 
Below, we discuss the results and our observations of these three categories.

\begin{table}[!htbp]
\caption{Categories of Developers' Feedback to CodeRabbit's Agentic Code Reviews\\and Mann-Kendall Trend Test of Distribution}
\label{tab:feedback_categories}
\centering
\resizebox{0.7\columnwidth}{!}{%
\huge
\sffamily 
\begin{tabular}{|lrrrrrr|}
\hline
\rowcolor[HTML]{EAEAEA} 
\textbf{Feedback Categories} & \multicolumn{1}{c}{\textbf{(n)}} & \multicolumn{1}{c}{\textbf{(\%)}} & \multicolumn{1}{c}{\textbf{($\tau$)}} & \multicolumn{1}{c}{\textbf{($p$)}} & \multicolumn{1}{c|}{\textbf{($\beta$)}} \\ \hline
\rowcolor[HTML]{FFFFFF} 
\multicolumn{1}{|l|}{\cellcolor[HTML]{FFFFFF}\begin{tabular}[c]{@{}l@{}}\textbf{Accepted} -  The developer found the code review useful and \\chose to incorporate the suggestion into a code change\end{tabular}} & 11,297 & 36.4 & \multicolumn{1}{|r}{0.457} & 0.002 & \multicolumn{1}{r|}{0.005} \\
\rowcolor[HTML]{EFEFEF} 
\multicolumn{1}{|l|}{\cellcolor[HTML]{EFEFEF}\begin{tabular}[c]{@{}l@{}}\textbf{Triggered Discussion}- The code review did not induce direct \\action, instead it triggered further discussion between human \\code reviewers and developers\end{tabular}} & 2,276 & 7.3 & \multicolumn{1}{|r}{-0.130} & 0.385 & \multicolumn{1}{r|}{0.000} \\ 
\multicolumn{1}{|l|}{\cellcolor[HTML]{FFFFFF}\begin{tabular}[c]{@{}l@{}}\textbf{Rejected} - The code review was either rejected due to the \\developer's own preferences and intentions or due to the \\code review itself being incorrect \end{tabular}} & 17,500 & 56.3 & \multicolumn{1}{|r}{-0.478} & 0.001 & \multicolumn{1}{r|}{-0.005} \\\hline
\rowcolor[HTML]{EFEFEF} 
\multicolumn{1}{|l}{\cellcolor[HTML]{EFEFEF}\textbf{Total Code Reviews: 31,073}} & & & & & \multicolumn{1}{l|}{}\\ \hline
\multicolumn{6}{l}{n = sample size, $\tau$ = Kendall's Tau, $p$ = p-value, $\beta$ = Sen's Slope}
\end{tabular}
}
\end{table}

The accepted suggestion provides a positive signal of using agentic code review tools in practice, indicating that the developer found the agentic code review to be helpful.
We observe that developers accepted the suggestion by incorporating the feedback into a code change, such as by (1) submitting a new commit, e.g., \textit{``fixed in new commit''}, (2) opening a corresponding PR, e.g., \textit{``Will be resolved by \#2981''}, or (3) creating an issue ticket, e.g., \textit{``I opened issue \#9080''}.
On the other hand, some agentic code reviews were not acted upon directly.
Rather, human developers and reviewers used them to trigger a discussion.
We observed that a human reviewer often initiates the discussion when they perceive that the agentic code review may be helpful, but require further confirmation from the developer who authored the PR or the wider development team.
For example, \textit{``@[developer name] this seems like a good point. why do we ...?''}, \textit{``@[developer name] do you think this comment needs to be addressed? Or do we desire the ... ''}.
Similarly, the developer can be uncertain about the validity of the agentic code review and seek input from other team members who possess a more comprehensive understanding of the overall codebase. 
For example, \textit{``Checking this with backend team, since it could be that ....''}, \textit{``Hey @[Reviewer name] what about this issue? do we need any changes?''}.
Although these scenarios do not directly result in code changes like accepted suggestions, these code reviews are still valuable because they facilitate knowledge transfer, similar to the traditional code review process conducted by human reviewers~\cite{6606617}.

When analysing trends over time, Table~\ref{tab:feedback_categories} shows that changes in both accepted and rejected suggestions were statistically significant (p $\leq$ 0.05). On average, the proportion of accepted suggestions increased by 0.5\% per month (approx. 12.5\% over 25 months), while the proportion of rejected suggestions showed a corresponding decline. 
These results suggest that CodeRabbit’s agentic code review system has become increasingly useful over time.

\summary{
Developers accepted 36.4\% of the agentic code reviews, leading to  new commits, PRs, or issue tickets. 
On the other hand, 7.3\% of agentic code reviews triggered human discussions, facilitating knowledge transfer among developers.
Over time, we found that the proportion of accepted reviews has increased by 12.5\% since release.
These results highlight the potential benefits and improvements of CodeRabbit's agentic code reviews.
}

\begin{table}[!htbp]
\caption{Rejection Reasons for Agentic Code Reviews\\and Mann-Kendall Trend Test of Distribution}
\label{tab:rejection_reasons}
\centering
\resizebox{0.7\columnwidth}{!}{%
\huge
\sffamily 
\begin{tabular}{|lrrrrrr|}
\hline
\rowcolor[HTML]{EAEAEA} 
\textbf{Rejection Reasons} & \multicolumn{1}{c}{\textbf{(n)}} & \multicolumn{1}{c}{\textbf{(\%)}} & \multicolumn{1}{c}{\textbf{($\tau$)}} & \multicolumn{1}{c}{\textbf{($p$)}} & \multicolumn{1}{c|}{\textbf{($\beta$)}} \\ \hline

\rowcolor[HTML]{EAEAEA} 
\multicolumn{1}{|l|}{\cellcolor[HTML]{EAEAEA}\textit{\textbf{Misalignment with Coding Practices}}} & \textbf{7,345} & \textbf{42.0} & \multicolumn{1}{|r}{0.304} & 0.040 &
 \multicolumn{1}{r|}{0.003} \\ \hline

\rowcolor[HTML]{FFFFFF} 
\multicolumn{1}{|l|}{\cellcolor[HTML]{FFFFFF}\begin{tabular}[c]{@{}l@{}}\textbf{Intended Design Trade-off} - The agentic code review was suggesting \\ to improve a suboptimal code implementation that was intentionally \\ chosen by the developer\end{tabular}} & 4,141 & 23.7 & \multicolumn{1}{|r}{0.362} & 0.014 & \multicolumn{1}{r|}{0.003} \\

\rowcolor[HTML]{EFEFEF} 
\multicolumn{1}{|l|}{\cellcolor[HTML]{EFEFEF}\begin{tabular}[c]{@{}l@{}}\textbf{Developer Preference} - The agentic code review was suggesting a \\ code improvement that diverges from the developer's personal \\ coding preferences\end{tabular}} & 3,204 & 18.3 & \multicolumn{1}{|r}{0.058} & 0.710 & \multicolumn{1}{r|}{0.001} \\ \hline

\rowcolor[HTML]{EAEAEA} 
\multicolumn{1}{|l|}{\cellcolor[HTML]{EAEAEA}\textit{\textbf{Invalid Suggestion}}} & \textbf{10,155} & \textbf{58.0} & \multicolumn{1}{|r}{-0.304} & 0.040 &
 \multicolumn{1}{r|}{-0.003} \\ \hline

\rowcolor[HTML]{FFFFFF} 
\multicolumn{1}{|l|}{\cellcolor[HTML]{FFFFFF}\begin{tabular}[c]{@{}l@{}}\textbf{False Positive Issue} - The agentic code review was suggesting to fix \\ an issue with the code that was actually misidentified and did not \\ constitute an actual defect\end{tabular}} & 7,570 & 43.3 & \multicolumn{1}{|r}{-0.159} & 0.286 & \multicolumn{1}{r|}{-0.003} \\

\rowcolor[HTML]{EFEFEF} 
\multicolumn{1}{|l|}{\cellcolor[HTML]{EFEFEF}\begin{tabular}[c]{@{}l@{}}\textbf{Redundant Suggestion} - The agentic code review was either suggesting \\ a code improvement that had already been implemented or identifying \\ an issue that was already being addressed\end{tabular}} & 822 & 4.7 & \multicolumn{1}{|r}{-0.123} & 0.413 & \multicolumn{1}{r|}{0.000} \\

\rowcolor[HTML]{FFFFFF} 
\multicolumn{1}{|l|}{\cellcolor[HTML]{FFFFFF}\begin{tabular}[c]{@{}l@{}}\textbf{Out of Scope} - The agentic code review was suggesting a code \\ improvement that was beyond the objective of the pull request that \\ was undergoing the code review\end{tabular}} & 1,763 & 10.0 & \multicolumn{1}{|r}{0.283} & 0.056 & \multicolumn{1}{r|}{0.001} \\ \hline

\rowcolor[HTML]{EFEFEF} 
\multicolumn{1}{|l}{\cellcolor[HTML]{EFEFEF}\textbf{Total Rejected Code Review Comments: 17,500}} & & & & & \multicolumn{1}{r|}{} \\ \hline
\multicolumn{6}{l}{n = sample size, $\tau$ = Kendall's Tau, $p$ = p-value, $\beta$ = Sen's Slope}
\end{tabular}
}
\end{table}

Despite finding promising benefits of agentic code reviews, the 56.3\% rejection rate highlights potential concerns associated with adopting such tools. 
Thus, we further conducted open coding on the reasons behind the 17,500	 rejected suggestions.
Table~\ref{tab:rejection_reasons} shows the rejection reasons, their descriptions, as well as their distribution. All percentages are reported based on the total rejected suggestions.
We observed two main scenarios: (1) agentic code reviews that were valid but misaligned with developers’ coding practices (42.0\%), and (2) agentic code reviews that were invalid (58.0\%).

For the agentic code reviews that were valid but misaligned with developers’ coding practices, we observed that developers disagree with CodeRabbit's suggestions due to intended design trade-offs (23.7\%) or personal coding preferences (18.3\%).
For intended design trade-offs, we observed that developers intentionally introduced a suboptimal implementation which was raised as a concern in the code review.
In these cases, developers typically focused on temporary workarounds, prototypes, test mocks, or simple solutions that were easier and faster to implement.
For example, Figure~\ref{fig:reject_misaligned_examples_intended} shows the agentic code review  recommending to add comprehensive error handling to the newly introduced function. However, the developer was still in the prototyping phase, focusing on minimally viable code and intentionally avoiding additional complexity in favour of rapid software development.
Similarly, for developer preference shown in Figure~\ref{fig:reject_misaligned_examples_preference}, CodeRabbit recommended a more concise implementation of similar assertions using mapping and loops instead of verbose boilerplate code, citing improved readability as the primary motivation.
In response, the developer rejected the suggestion, as they considered their implementation to be sufficiently readable.
These findings suggest that even when CodeRabbit's agentic code reviews are technically valid, their usefulness depends on how well they accommodate developers’ intentions and preferences under the respective context.

For the agentic code reviews that were deemed invalid, we found that the reviews often raised false positive issues (43.3\%), offered redundant suggestions (4.7\%), or surfaced concerns that were out of scope for the pull request (10.0\%). 
For the case of false positives, the agent flagged non‑existent issues and proposed correspondingly misguided fixes. 
For example, as shown in Figure~\ref{fig:reject_invalid_examples_false}, the agent warned of a potential infinite loop in the pagination logic and recommended revising the control flow to prevent unbounded execution. The developer clarified that the server handled this logic, rendering the concern invalid.
For redundant suggestions, we observed cases where the agent proposed improvements that were already implemented or had been previously addressed by the developers. 
In Figure~\ref{fig:reject_invalid_examples_redundant}, the agent noted that several Jupyter images were pinned to an outdated library version due to dependency constraints and advised creating an issue ticket. 
The developer replied that a ticket for this matter already existed and reminded the agent that they had raised this same point before.
Lastly, for out‑of‑scope suggestions, the agent surfaced concerns unrelated to the PR’s purpose. 
In Figure~\ref{fig:reject_invalid_examples_out}, the agent warned of possible incompatibilities between inline React styles and advanced selectors.
The developer noted that this suggestion is out of scope, as the PR focused solely on refactoring the Budget Name component.
These findings suggest that agentic code reviews can misinterpret the code implementation, overlook existing information, or drift beyond the intended scope of changes, highlighting the need to enhance the agents’ understanding of both the software system’s implementation and its development process.

When analysing trends over time, Table~\ref{tab:rejection_reasons} shows that misalignment with coding practices has exhibited a statistically significant increase of 0.3\% per month (approx. 7.5\% over 25 months), while the proportion of invalid suggestions has shown a corresponding decrease. 
At the subcategory level, no statistically significant trends were observed, with the exception of a steady increase in intended design trade-offs. 
This indicates that while CodeRabbit is becoming more technically proficient over time, its ability to adapt to developer-specific contexts and intents remains limited.

\begin{figure}[!htbp]
\begin{tcolorbox}[colback=gray!5!white, colframe=lightgray, coltitle=black, fonttitle=\sffamily\bfseries\footnotesize, fontupper=\sffamily\footnotesize, title=Intended Design Trade-off]

\textbf{Simplified Diff:}
\begin{lstlisting}[basicstyle=\ttfamily\footnotesize]
19 +function getWahlvorstand(wahlbezirkID: string): Promise<Wahlvorstand> {
20 +  return wahlvorstandControllerApi
21 +    .getWahlvorstand(wahlbezirkID)
22 +    .then((response) => toModel(response.data));
\end{lstlisting}
\textbf{Agentic Review:}  
{\footnotesize Refactor suggestion. Add error handling to getWahlvorstand function. The function doesn't handle potential errors from the API call, which could lead to unhandled promise rejections.}
\begin{lstlisting}[basicstyle=\ttfamily\footnotesize]
22 -   .then((response) => toModel(response.data));
23 +   .then((response) => toModel(response.data))
24 +   .catch((error) => {
25 +     console.error("Failed to fetch Wahlvorstand:", error);
26 +     throw error;
27 +   });
\end{lstlisting} 

\textbf{Human Developer:} \\
No error handling intended during this prototype phase.
\end{tcolorbox}
\caption{\normalsize Example of an Agentic Code Review Rejected Due to Intended Design Trade-off}
\label{fig:reject_misaligned_examples_intended}
\Description{misalignment}
\end{figure}

\begin{figure}[!htbp]
\begin{tcolorbox}[colback=gray!5!white, colframe=lightgray, coltitle=black, fonttitle=\bfseries\sffamily\footnotesize , fontupper=\sffamily\footnotesize, title=Developer Preference]
\textbf{Simplified Diff:}
\begin{lstlisting}[basicstyle=\ttfamily\footnotesize]
72 -self.assertNoHoliday(Portugal(subdiv="01"), "1910-05-12")
73 -self.assertNoHoliday(Portugal(subdiv="02"), "1910-05-05")
74 -self.assertNoHoliday(Portugal(subdiv="03"), "1910-06-24")
...
72 +self.assertNoSubdiv01Holiday("1910-05-12")
73 +self.assertNoSubdiv02Holiday("1910-05-05")
74 +self.assertNoSubdiv03Holiday("1910-06-24"
...
\end{lstlisting}
\textbf{Agentic Review:} \\
Nitpick (assertive). Readable per‑subdivision checks. Explicit assertions are fine. If desired, we can condense via a mapping + loop without losing clarity. I can provide a compact loop-based variant on request. Also applies to: 94-114.\\
\textbf{Human Developer:} \\
I'd argue that the existing implementation is still human-readable.
\end{tcolorbox}
\caption{\normalsize Example of an Agentic Code Review Rejected Due to Developer Preference}
\label{fig:reject_misaligned_examples_preference}
\Description{misalignment}
\end{figure}

\begin{figure}[!htbp]
\begin{tcolorbox}[colback=gray!5!white, colframe=lightgray, coltitle=black, fonttitle=\bfseries\sffamily, fontupper=\sffamily\footnotesize, title=False Positive Issue ]
\textbf{Simplified Diff:}
\begin{lstlisting}[basicstyle=\ttfamily\footnotesize]
66 +questionsLoaded = computed(() => this.allLoadedQuestions().length > 0);
67 +nextPage = computed(() => (this.currentIndex() + 2) % this.size === 0 && this.hasNext());
\end{lstlisting}
\textbf{Agentic Review:} \\
Potential issue. Critical. Fix infinite auto-pagination loop; gate requests and handle errors. nextPage ignores page(), so the effect re-triggers loadNextPage() indefinitely (page++ doesn’t affect nextPage). Add gating and compute against page; also set/reset a loading flag and handle HTTP errors. [...] \\
\textbf{Human Developer:} \\
The server determines if there is a next page and sets the header accordingly. Only if hasNext is true (set by the server) the next page can get loaded. So this is not a problem.
\end{tcolorbox}

\caption{\normalsize Example of an Agentic Code Review Rejected Due to False Positive Issue}
\label{fig:reject_invalid_examples_false}
\Description{invalid}
\end{figure}

\begin{figure}[!htbp]
\begin{tcolorbox}[colback=gray!5!white, colframe=lightgray, coltitle=black, fonttitle=\bfseries\sffamily, fontupper=\sffamily\footnotesize, title=Redundant Suggestion]
\textbf{Simplified Diff:}
\begin{lstlisting}[basicstyle=\ttfamily\footnotesize]
2092 +version = "4.2.4"
2093 +sdist = { url = "https://files.pythonhosted.org/packages/.../odh_elyra-4.2.4.tar.gz", upload-time = 2025-09-11T18:09:43Z, size = 2155769, hashes = { sha256 = "..." } }
2094 +wheels = [{ url = "https://files.pythonhosted.org/packages/.../odh_elyra-4.2.4-py3-none-any.whl", upload-time = 2025-09-11T18:09:41Z, size = 4317667, hashes = { sha256 = "..." }}]
\end{lstlisting}

\textbf{Agentic Review:} \\
Confirmed: odh-elyra 4.2.4 pulls appengine-python-standard (1.1.10) and keeps urllib3 at 1.26.x (\textless2) in Jupyter images.\\
\textbf{Human Developer:} \\
There already is urllib3 tracking issue. Add a comment there. I asked you in the past to create issue for this in elyra-ai/elyra repository.
\end{tcolorbox}
\caption{\normalsize Example of an Agentic Code Review Rejected Due to Redundant Suggestion}
\label{fig:reject_invalid_examples_redundant}
\Description{invalid}
\end{figure}

\begin{figure}[!htbp]
\begin{tcolorbox}[colback=gray!5!white, colframe=lightgray, coltitle=black, fonttitle=\sffamily\bfseries, fontupper=\footnotesize, title=Out of Scope]
\textbf{PR Title:} \\
Update Sidebar - Refactor the Budget Name component \#3593

\textbf{PR Description:} \\
Continuation of splitting \#3457 into several PR's as requested by @Human Developer.
This PR removes "Rename budget" from the menu and replaces it with a pencil icon next to the budget name. This updates the sidebar to the same look/feel as renaming an account from the account page. [...]

\textbf{Simplified Diff:}
\begin{lstlisting}[basicstyle=\ttfamily\footnotesize]
65 +'& .hover-visible': {
66 + opacity: 0,
67 + transition: 'opacity .25s',
68 +},
69 +'&:hover .hover-visible': {
70 + opacity: 1,
71 +},
\end{lstlisting}
\textbf{Agentic Review:} \\
Potential issue. CSS-in-JS syntax might not work as expected. [...] 
Consider using a CSS-in-JS solution like styled-components or Emotion, [...] \\
\textbf{Human Developer:} \\
This is beyond the scope of this PR.
\end{tcolorbox}
\caption{\normalsize Example of an Agentic Code Review Rejected Due to Out of Scope}
\label{fig:reject_invalid_examples_out}
\Description{invalid}
\end{figure}

\summary{
For the remaining 56.3\% of code reviews that developers rejected, the primary reasons were misalignment with coding practices and invalid suggestions. 
False positive issues were the most common cause (43.3\%), followed by intentional design trade‑offs (23.7\%).
While the proportion of invalid suggestions have decreased by 7.5\% since release, misalignment with coding practices has correspondingly increased, suggesting that CodeRabbit is becoming more technically proficient but remain unable to adapt to developer-specific contexts and intents.
}

\subsection*{(RQ2) \rqtwo}
\begin{figure}[!htbp]
  \centering
  \includegraphics[width=0.6\columnwidth]{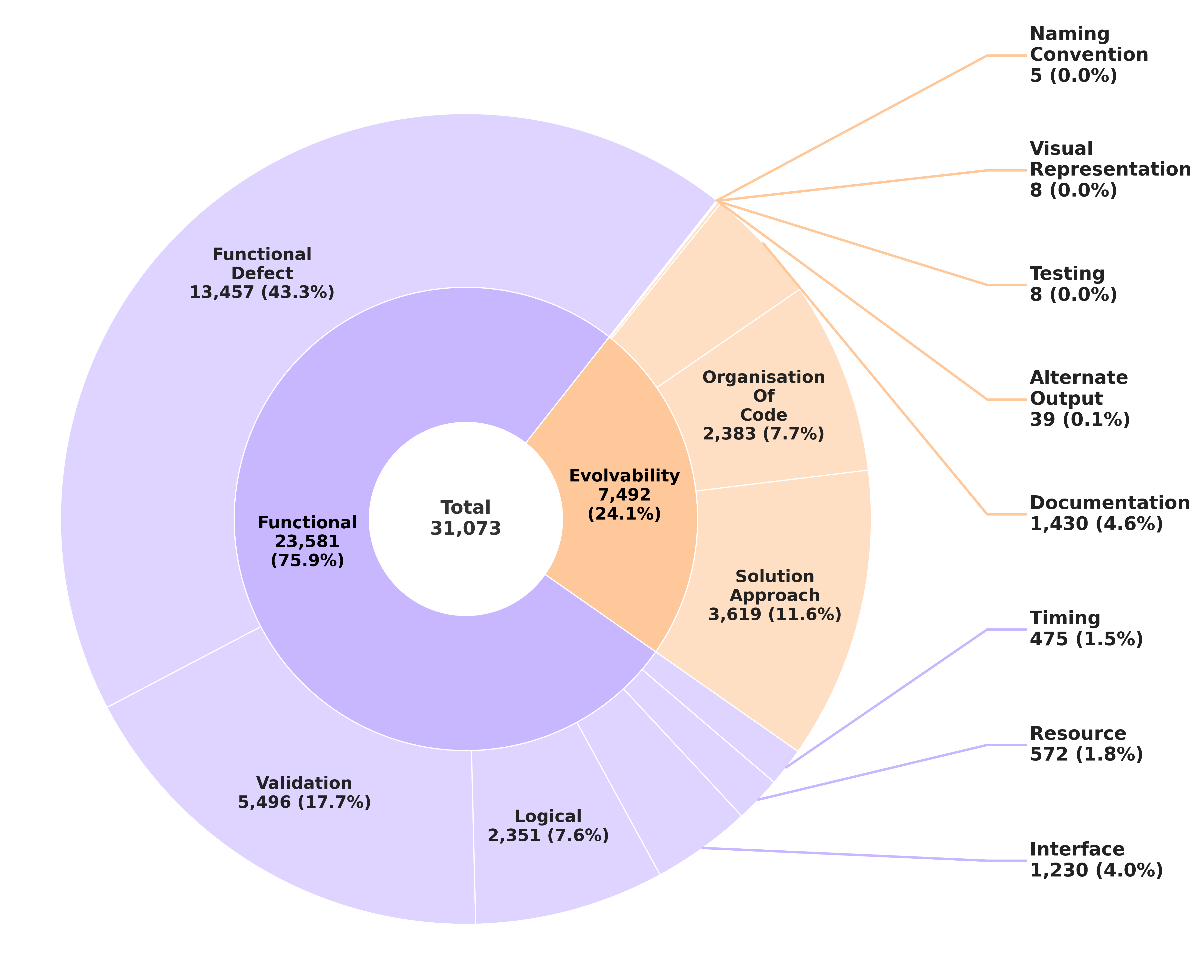}
  \caption{Distribution of Comment Types Provided in Agentic Code Reviews}
  \label{fig:code_review_categories}
  \Description{categories}
\end{figure}

\textbf{Approach.}
To address this question, we manually classified the 297 agentic code reviews into categories established by prior research~\cite{turzo2024makes,mika,beller}.
At a high level, code reviews can be categorised into two main types: functional and evolvability.
Code reviews targeting functional aspects focus on immediate defects that can lead to system failures at execution time.
In contrast, evolvability-related reviews address longer-term concerns, including compliance, maintainability, and understandability of the code.
These two main categories can be further separated into the subcategories shown in Figure~\ref{fig:code_review_categories}.
To that understanding of the annotation criteria, the first two authors independently annotated 30 samples, achieving substantial inter-rater agreement (Cohen’s $\kappa$ = 0.85). 
Discrepancies were then discussed and resolved between the two annotators.
When agreement could not be reached, the third author examined the discrepancies to achieve a consensus. 
Given the substantial agreement achieved in the first round, the first author continued to annotate the remaining samples based on this shared understanding.
All annotations were then validated by the second author for correctness.
Finally, we evaluated GPT-5.1's agreement with our manually derived labels when classifying the same samples.
Achieving substantial agreement (Cohen's $\kappa = 0.78$), the LLM was subsequently used to annotate the remaining 30,776 agentic code reviews.
The prompt that we utilised is shown in Figure~\ref{fig:cr_categorisation}.
We conduct the time series analysis at monthly intervals, reporting Kendall's Tau, the p-value, and Sen's slope from the Mann-Kendall trend test.

\begin{figure}[!htbp]
\begin{tcolorbox}[colback=gray!5!white, colframe=lightgray, coltitle=black, fonttitle=\bfseries\sffamily\footnotesize , fontupper=\sffamily\footnotesize, title=Code Review Comment Categorisation Prompt]

You are an expert software engineering researcher analysing human-AI interactions. 
Label this agent-generated code review by CodeRabbit AI into exactly one of the 13 predefined categories. \\

The category should be one of the following 13 types. \\

\textbf{1) Functional Defect:} A functionality is missing or implemented incorrectly, which often requires additional code or larger modification. \\
\textbf{2) Validation:} Issues with detecting an invalid value and issues related to data sanitisation. \\
\textbf{3) Logical:} Issues with comparison operations, control flow, computations and other types of logical errors. \\
\textbf{4) Interface:} Issues when interacting with other parts of the software, e.g., existing code library, hardware device, database, OS. \\
\textbf{5) Resource:} Issues with the initialisation, manipulation and release of variables, memory, files and database. \\
\textbf{6) Timing:} Issues with incorrect thread synchronisation in shared resource settings. \\
\textbf{7) Solution Approach:} Suggestions for alternate implementations, e.g., algorithms, data structures. \\
\textbf{8) Documentation:} Suggestions to improve code comments or documentation. \\
\textbf{9) Organisation of Code:} Suggestions for structural refactoring, e.g., collapse hierarchy, extract super class, inline function. \\
\textbf{10) Alternate Output:} Suggestions for improving error messages, toast messages, alerts and the returned values of a function. \\
\textbf{11) Naming Convention:} Suggestions for renaming software elements to comply with conventions. \\
\textbf{12) Visual Representation:} Suggestions for improving code readability, e.g., removing white spaces, blank lines, indentation. \\
\textbf{13) Testing:} Suggestions for adding, modifying, or improving tests, test coverage, assertions, mocks, fixtures, and other testing-related code. \\

{[}CODE\_REVIEW{]}\\
\{code\_review\}\\
{[}\textbackslash CODE\_REVIEW{]}\\

Directly answer with one of the 1-13 symbols:
\end{tcolorbox}

\caption{\normalsize Code Review Comment Categorisation Prompt}
\label{fig:cr_categorisation}
\Description{cr_categorisation}
\end{figure}

\textbf{Results.} Figure~\ref{fig:code_review_categories} shows the distribution of the comment types generated by CodeRabbit.
All percentages are reported based on the overall number of code reviews.
We find that the 75.9\% of the code reviews were related to the functional aspects of the code, whilst 24.1\% of them were related to evolvability.
Within the functional-related reviews, the most frequent type was functional defects (43.3\%), whilst the other five sub-types were far lower with a proportion of 1.5\% - 17.7\%.
For evolvability-related reviews, the most common types were solution approach (11.6\%) and organisation of code (7.7\%), whilst the other sub-types in this category were rare ($\leq$ 4.6\%).
These findings indicate that although CodeRabbit can provide diverse feedback, it has a pronounced focus on identifying functional defects.

When analysing trends over time, Table~\ref{tab:category_rejection_rate} shows a statistically significant increase of code reviews related to functional issues at 0.7\% per month (approx. 17.5\% over 25 months), with a corresponding decrease in those related to evolvability.
In terms of specific aspects of functionality, we find that reviews related to functional defect and interface have been steadily increasing at 0.9\% and 0.2\% per month, while reviews related to logical issues has been decreasing at 0.3\% per month.
In terms of specific aspects of evolvability, we find that code reviews related to documentation and organisation of code have been decreasing at 0.2\% and 0.5\%, respectively.
This indicates that CodeRabbit is increasingly 
prioritising technical concerns related to functional correctness over those concerned with the long-term maintainability of code.

Table~\ref{tab:category_rejection_rate} shows the distribution of developer feedback across the code review categories.
All percentages are reported based on the total number of code reviews in the respective category.
Overall, functional-related and
evolvability-related reviews have relatively similar rejection rates,
i.e., 56.2\% and 56.7\%, respectively.
Nonetheless, we found that functional code reviews were rejected as invalid more often than evolvability‑related reviews (35.1\% vs. 25.1\%), suggesting that the agent struggles more with accurately identifying functional defects than with evaluating evolvability‑related issues. 

Functional defects were the most common sub-type that CodeRabbit focused on, yet they yielded a 56.5\% rejection rate, with 35.8\% of suggestions flagged as invalid by the developer. 
Other, less frequent functional reviews saw similarly high rejection rates (48.6\%–57.8\%), largely because they were also deemed invalid (30.8\%–40.8\%). 
These findings highlight that while CodeRabbit actively targets functional issues, it struggles to produce reliable feedback in these areas.
Although certain evolvability-related reviews achieved higher acceptance rates (60.0\%–87.5\%), these were largely confined to long-tail categories, such as naming convention, visual representation, and testing.
For the rest of the categories with high representation e.g., solution approach, documentation, and organisation of code, they exhibited a similar rejection rate to code reviews focused on functional issues (54.2\%-59.4\%).
In contrast, however, these were mostly comprised of misalignments with coding practices (25.8\%-34.7\%), rather than being invalid suggestions that were hallucinated (23.5\%-28.7\%).
These findings indicate that despite offering technically valid suggestions for more optimised implementations, cleaner refactoring, and enhanced documentation, CodeRabbit's agentic reviews are frequently dismissed due to the developers intentionally retaining suboptimal code or simply preferring their own approach.

\begin{table}[!htbp]
\caption{Developer Responses by Code Review Category\\and Mann-Kendall Trend Test of Distribution}
\label{tab:category_rejection_rate}
\resizebox{0.7\columnwidth}{!}{%
\Large
\sffamily 
\begin{tabular}{|lrrrrrrrrrr|}
\hline
\multicolumn{1}{|l|}{\cellcolor[HTML]{EFEFEF}\textbf{}} & \multicolumn{1}{c|}{\cellcolor[HTML]{EFEFEF}\textbf{\begin{tabular}[c]{@{}c@{}}Accepted\\ (\%)\end{tabular}}} & \multicolumn{1}{c|}{\cellcolor[HTML]{EFEFEF}\textbf{\begin{tabular}[c]{@{}c@{}}Triggered \\ Discussion\\ (\%)\end{tabular}}} & \multicolumn{1}{c|}{\cellcolor[HTML]{EFEFEF}\textbf{\begin{tabular}[c]{@{}c@{}}Overall \\ Rejected\\ (\%)\end{tabular}}} & \multicolumn{1}{c|}{\cellcolor[HTML]{EFEFEF}\textbf{\begin{tabular}[c]{@{}c@{}}Misalign. \\ w/ Coding \\ Prac. (\%)\end{tabular}}} & \multicolumn{1}{c|}{\cellcolor[HTML]{EFEFEF}\textbf{\begin{tabular}[c]{@{}c@{}}Invalid \\ Suggest.\\ (\%)\end{tabular}}} & \multicolumn{1}{c|}{\cellcolor[HTML]{EFEFEF}\textbf{\begin{tabular}[c]{@{}c@{}}Total \\ Number \\(\textbf{n})\end{tabular}}} & \multicolumn{1}{c|}{\cellcolor[HTML]{EFEFEF}\textbf{($\tau$)}} & \multicolumn{1}{c|}{\cellcolor[HTML]{EFEFEF}\textbf{($p$)}} & \multicolumn{1}{c|}{\cellcolor[HTML]{EFEFEF}\textbf{($\beta$)}} \\ \hline

\rowcolor[HTML]{EFEFEF} 
\multicolumn{10}{|l|}{\cellcolor[HTML]{EFEFEF}\textbf{Main Category}} \\ \hline

\rowcolor[HTML]{FFFFFF} 
\multicolumn{1}{|l|}{\cellcolor[HTML]{C8B6FF}Functional} & \multicolumn{1}{r|}{\cellcolor[HTML]{FFFFFF}36.5} & \multicolumn{1}{r|}{\cellcolor[HTML]{FFFFFF}7.3} & \multicolumn{1}{r|}{\cellcolor[HTML]{FFFFFF}56.2} & \multicolumn{1}{r|}{\cellcolor[HTML]{FFFFFF}21.1} & \multicolumn{1}{r|}{\cellcolor[HTML]{FFFFFF}35.1} & \multicolumn{1}{r|}{\cellcolor[HTML]{FFFFFF}23,581} & \multicolumn{1}{r|}{\cellcolor[HTML]{FFFFFF}0.5} & \multicolumn{1}{r|}{\cellcolor[HTML]{FFFFFF}0.001} & \multicolumn{1}{r|}{\cellcolor[HTML]{FFFFFF}0.007} \\

\rowcolor[HTML]{EFEFEF} 
\multicolumn{1}{|l|}{\cellcolor[HTML]{FEC89A}Evolvability} & \multicolumn{1}{r|}{\cellcolor[HTML]{EFEFEF}35.8} & \multicolumn{1}{r|}{\cellcolor[HTML]{EFEFEF}7.5} & \multicolumn{1}{r|}{\cellcolor[HTML]{EFEFEF}56.7} & \multicolumn{1}{r|}{\cellcolor[HTML]{EFEFEF}31.6} & \multicolumn{1}{r|}{\cellcolor[HTML]{EFEFEF}25.1} & \multicolumn{1}{r|}{\cellcolor[HTML]{EFEFEF}7,492} & \multicolumn{1}{r|}{\cellcolor[HTML]{EFEFEF}-0.5} & \multicolumn{1}{r|}{\cellcolor[HTML]{EFEFEF}0.001} & \multicolumn{1}{r|}{\cellcolor[HTML]{EFEFEF}-0.007} \\ \hline

\rowcolor[HTML]{EFEFEF} 
\multicolumn{10}{|l|}{\cellcolor[HTML]{EFEFEF}\textbf{Subcategory}} \\ \hline

\rowcolor[HTML]{FFFFFF} 
\multicolumn{1}{|l|}{\cellcolor[HTML]{C8B6FF}Functional Defect} & \multicolumn{1}{r|}{\cellcolor[HTML]{FFFFFF}36.2} & \multicolumn{1}{r|}{\cellcolor[HTML]{FFFFFF}7.3} & \multicolumn{1}{r|}{\cellcolor[HTML]{FFFFFF}56.5} & \multicolumn{1}{r|}{\cellcolor[HTML]{FFFFFF}20.7} & \multicolumn{1}{r|}{\cellcolor[HTML]{FFFFFF}35.8} & \multicolumn{1}{r|}{\cellcolor[HTML]{FFFFFF}13,457} & \multicolumn{1}{r|}{\cellcolor[HTML]{FFFFFF}0.471} & \multicolumn{1}{r|}{\cellcolor[HTML]{FFFFFF}0.001} & \multicolumn{1}{r|}{\cellcolor[HTML]{FFFFFF}0.009} \\

\rowcolor[HTML]{EFEFEF} 
\multicolumn{1}{|l|}{\cellcolor[HTML]{C8B6FF}Validation} & \multicolumn{1}{r|}{\cellcolor[HTML]{EFEFEF}36.4} & \multicolumn{1}{r|}{\cellcolor[HTML]{EFEFEF}6.0} & \multicolumn{1}{r|}{\cellcolor[HTML]{EFEFEF}57.6} & \multicolumn{1}{r|}{\cellcolor[HTML]{EFEFEF}23.8} & \multicolumn{1}{r|}{\cellcolor[HTML]{EFEFEF}33.8} & \multicolumn{1}{r|}{\cellcolor[HTML]{EFEFEF}5,496} & \multicolumn{1}{r|}{\cellcolor[HTML]{EFEFEF}-0.058} & \multicolumn{1}{r|}{\cellcolor[HTML]{EFEFEF}0.710} & \multicolumn{1}{r|}{\cellcolor[HTML]{EFEFEF}-0.001} \\

\rowcolor[HTML]{FFFFFF} 
\multicolumn{1}{|l|}{\cellcolor[HTML]{C8B6FF}Logical} & \multicolumn{1}{r|}{\cellcolor[HTML]{FFFFFF}40.4} & \multicolumn{1}{r|}{\cellcolor[HTML]{FFFFFF}7.2} & \multicolumn{1}{r|}{\cellcolor[HTML]{FFFFFF}52.4} & \multicolumn{1}{r|}{\cellcolor[HTML]{FFFFFF}19.9} & \multicolumn{1}{r|}{\cellcolor[HTML]{FFFFFF}32.5} & \multicolumn{1}{r|}{\cellcolor[HTML]{FFFFFF}2,351} & \multicolumn{1}{r|}{\cellcolor[HTML]{FFFFFF}-0.514} & \multicolumn{1}{r|}{\cellcolor[HTML]{FFFFFF}0.000} & \multicolumn{1}{r|}{\cellcolor[HTML]{FFFFFF}-0.003} \\

\rowcolor[HTML]{EFEFEF} 
\multicolumn{1}{|l|}{\cellcolor[HTML]{C8B6FF}Interface} & \multicolumn{1}{r|}{\cellcolor[HTML]{EFEFEF}33.6} & \multicolumn{1}{r|}{\cellcolor[HTML]{EFEFEF}8.6} & \multicolumn{1}{r|}{\cellcolor[HTML]{EFEFEF}57.8} & \multicolumn{1}{r|}{\cellcolor[HTML]{EFEFEF}17.0} & \multicolumn{1}{r|}{\cellcolor[HTML]{EFEFEF}40.8} & \multicolumn{1}{r|}{\cellcolor[HTML]{EFEFEF}1,230} & \multicolumn{1}{r|}{\cellcolor[HTML]{EFEFEF}0.630} & \multicolumn{1}{r|}{\cellcolor[HTML]{EFEFEF}0.000} & \multicolumn{1}{r|}{\cellcolor[HTML]{EFEFEF}0.002} \\

\rowcolor[HTML]{FFFFFF} 
\multicolumn{1}{|l|}{\cellcolor[HTML]{C8B6FF}Resource} & \multicolumn{1}{r|}{\cellcolor[HTML]{FFFFFF}40.4} & \multicolumn{1}{r|}{\cellcolor[HTML]{FFFFFF}11.0} & \multicolumn{1}{r|}{\cellcolor[HTML]{FFFFFF}48.6} & \multicolumn{1}{r|}{\cellcolor[HTML]{FFFFFF}17.8} & \multicolumn{1}{r|}{\cellcolor[HTML]{FFFFFF}30.8} & \multicolumn{1}{r|}{\cellcolor[HTML]{FFFFFF}572} & \multicolumn{1}{r|}{\cellcolor[HTML]{FFFFFF}-0.109} & \multicolumn{1}{r|}{\cellcolor[HTML]{FFFFFF}0.472} & \multicolumn{1}{r|}{\cellcolor[HTML]{FFFFFF}0.000} \\

\rowcolor[HTML]{EFEFEF} 
\multicolumn{1}{|l|}{\cellcolor[HTML]{C8B6FF}Timing} & \multicolumn{1}{r|}{\cellcolor[HTML]{EFEFEF}32.6} & \multicolumn{1}{r|}{\cellcolor[HTML]{EFEFEF}12.2} & \multicolumn{1}{r|}{\cellcolor[HTML]{EFEFEF}55.2} & \multicolumn{1}{r|}{\cellcolor[HTML]{EFEFEF}23.6} & \multicolumn{1}{r|}{\cellcolor[HTML]{EFEFEF}31.6} & \multicolumn{1}{r|}{\cellcolor[HTML]{EFEFEF}475} & \multicolumn{1}{r|}{\cellcolor[HTML]{EFEFEF}0.130} & \multicolumn{1}{r|}{\cellcolor[HTML]{EFEFEF}0.385} & \multicolumn{1}{r|}{\cellcolor[HTML]{EFEFEF}0.000} \\

\rowcolor[HTML]{FFFFFF} 
\multicolumn{1}{|l|}{\cellcolor[HTML]{FEC89A}Solution Approach} & \multicolumn{1}{r|}{\cellcolor[HTML]{FFFFFF}32.0} & \multicolumn{1}{r|}{\cellcolor[HTML]{FFFFFF}8.6} & \multicolumn{1}{r|}{\cellcolor[HTML]{FFFFFF}59.4} & \multicolumn{1}{r|}{\cellcolor[HTML]{FFFFFF}34.7} & \multicolumn{1}{r|}{\cellcolor[HTML]{FFFFFF}24.6} & \multicolumn{1}{r|}{\cellcolor[HTML]{FFFFFF}3,619} & \multicolumn{1}{r|}{\cellcolor[HTML]{FFFFFF}-0.014} & \multicolumn{1}{r|}{\cellcolor[HTML]{FFFFFF}0.941} & \multicolumn{1}{r|}{\cellcolor[HTML]{FFFFFF}0.000} \\

\rowcolor[HTML]{EFEFEF} 
\multicolumn{1}{|l|}{\cellcolor[HTML]{FEC89A}Documentation} & \multicolumn{1}{r|}{\cellcolor[HTML]{EFEFEF}39.9} & \multicolumn{1}{r|}{\cellcolor[HTML]{EFEFEF}5.6} & \multicolumn{1}{r|}{\cellcolor[HTML]{EFEFEF}54.5} & \multicolumn{1}{r|}{\cellcolor[HTML]{EFEFEF}25.8} & \multicolumn{1}{r|}{\cellcolor[HTML]{EFEFEF}28.7} & \multicolumn{1}{r|}{\cellcolor[HTML]{EFEFEF}1,430} & \multicolumn{1}{r|}{\cellcolor[HTML]{EFEFEF}-0.428} & \multicolumn{1}{r|}{\cellcolor[HTML]{EFEFEF}0.004} & \multicolumn{1}{r|}{\cellcolor[HTML]{EFEFEF}-0.002} \\

\rowcolor[HTML]{FFFFFF} 
\multicolumn{1}{|l|}{\cellcolor[HTML]{FEC89A}Organisation of Code} & \multicolumn{1}{r|}{\cellcolor[HTML]{FFFFFF}38.7} & \multicolumn{1}{r|}{\cellcolor[HTML]{FFFFFF}7.1} & \multicolumn{1}{r|}{\cellcolor[HTML]{FFFFFF}54.2} & \multicolumn{1}{r|}{\cellcolor[HTML]{FFFFFF}30.7} & \multicolumn{1}{r|}{\cellcolor[HTML]{FFFFFF}23.5} & \multicolumn{1}{r|}{\cellcolor[HTML]{FFFFFF}2,383} & \multicolumn{1}{r|}{\cellcolor[HTML]{FFFFFF}-0.507} & \multicolumn{1}{r|}{\cellcolor[HTML]{FFFFFF}0.001} & \multicolumn{1}{r|}{\cellcolor[HTML]{FFFFFF}-0.005} \\

\rowcolor[HTML]{EFEFEF} 
\multicolumn{1}{|l|}{\cellcolor[HTML]{FEC89A}Alternate Output} & \multicolumn{1}{r|}{\cellcolor[HTML]{EFEFEF}43.6} & \multicolumn{1}{r|}{\cellcolor[HTML]{EFEFEF}2.6} & \multicolumn{1}{r|}{\cellcolor[HTML]{EFEFEF}53.8} & \multicolumn{1}{r|}{\cellcolor[HTML]{EFEFEF}12.8} & \multicolumn{1}{r|}{\cellcolor[HTML]{EFEFEF}41.0} & \multicolumn{1}{r|}{\cellcolor[HTML]{EFEFEF}39} & \multicolumn{1}{r|}{\cellcolor[HTML]{EFEFEF}0.029} & \multicolumn{1}{r|}{\cellcolor[HTML]{EFEFEF}0.859} & \multicolumn{1}{r|}{\cellcolor[HTML]{EFEFEF}0.000} \\

\rowcolor[HTML]{FFFFFF} 
\multicolumn{1}{|l|}{\cellcolor[HTML]{FEC89A}Naming Convention} & \multicolumn{1}{r|}{\cellcolor[HTML]{FFFFFF}60.0} & \multicolumn{1}{r|}{\cellcolor[HTML]{FFFFFF}0.0} & \multicolumn{1}{r|}{\cellcolor[HTML]{FFFFFF}40.0} & \multicolumn{1}{r|}{\cellcolor[HTML]{FFFFFF}20.0} & \multicolumn{1}{r|}{\cellcolor[HTML]{FFFFFF}20.0} & \multicolumn{1}{r|}{\cellcolor[HTML]{FFFFFF}5} & \multicolumn{1}{r|}{\cellcolor[HTML]{FFFFFF}0.091} & \multicolumn{1}{r|}{\cellcolor[HTML]{FFFFFF}0.399} & \multicolumn{1}{r|}{\cellcolor[HTML]{FFFFFF}0.000} \\

\rowcolor[HTML]{EFEFEF} 
\multicolumn{1}{|l|}{\cellcolor[HTML]{FEC89A}Visual Representation} & \multicolumn{1}{r|}{\cellcolor[HTML]{EFEFEF}62.5} & \multicolumn{1}{r|}{\cellcolor[HTML]{EFEFEF}0.0} & \multicolumn{1}{r|}{\cellcolor[HTML]{EFEFEF}37.5} & \multicolumn{1}{r|}{\cellcolor[HTML]{EFEFEF}37.5} & \multicolumn{1}{r|}{\cellcolor[HTML]{EFEFEF}0.0} & \multicolumn{1}{r|}{\cellcolor[HTML]{EFEFEF}8} & \multicolumn{1}{r|}{\cellcolor[HTML]{EFEFEF}0.188} & \multicolumn{1}{r|}{\cellcolor[HTML]{EFEFEF}0.130} & \multicolumn{1}{r|}{\cellcolor[HTML]{EFEFEF}0.000} \\

\rowcolor[HTML]{FFFFFF} 
\multicolumn{1}{|l|}{\cellcolor[HTML]{FEC89A}Testing} & \multicolumn{1}{r|}{\cellcolor[HTML]{FFFFFF}87.5} & \multicolumn{1}{r|}{\cellcolor[HTML]{FFFFFF}0.0} & \multicolumn{1}{r|}{\cellcolor[HTML]{FFFFFF}12.5} & \multicolumn{1}{r|}{\cellcolor[HTML]{FFFFFF}0.0} & \multicolumn{1}{r|}{\cellcolor[HTML]{FFFFFF}12.5} & \multicolumn{1}{r|}{\cellcolor[HTML]{FFFFFF}8} & \multicolumn{1}{r|}{\cellcolor[HTML]{FFFFFF}-0.011} & \multicolumn{1}{r|}{\cellcolor[HTML]{FFFFFF}0.948} & \multicolumn{1}{r|}{\cellcolor[HTML]{FFFFFF}0.000} \\ \hline

\multicolumn{10}{l}{\large Accepted + Triggered Discussion + Overall Rejected = 100\%, n = sample size, $\tau$ = Kendall's Tau, $p$ = p-value, $\beta$ = Sen's Slope}\\
\multicolumn{10}{l}{\large Misalignment w/ Coding Practices + Invalid Suggestion = Overall Rejected}\\
\multicolumn{10}{l}{}
\end{tabular}
}
\end{table}

\summary{
Most of CodeRabbit's reviews focus on functionality (75.9\%), with a growing emphasis over evolvability-related feedback (17.5\% since release), yet functional 
reviews face a high rejection rate (48.6\%--57.8\%), largely due to being 
invalid (30.8\%--40.8\%).
Similarly, evolvability‑related concerns tended to also have high rejection rates (54.2\%-59.4\%), however, they were mostly due to misalignment with the developers' coding practices (25.8\%-34.7\%).
While false positive suggestions for complex functional issues are problematic, valid suggestions for improving maintenance are still often rejected for failing to adequately conform to the developer's practices.
}

\subsection*{(RQ3) \rqthree}
\textbf{Approach.}
To address this question, we investigated methods for predicting whether an agentic code review produced by CodeRabbit will be rejected.
An effective predicter can be utilised as an inference time quality filter, preventing low quality reviews from reaching the developer.
Since the approach must be practical at inference time, we restrict our investigation to lightweight techniques with reasonable overhead in latency and computational cost.
This excludes agentic judge approaches capable of retrieving and reasoning over additional context, as their overhead is comparable to conducting an additional run of CodeRabbit's entire workflow.
Specifically, we evaluated three LLM-based approaches: direct prompting, fine-tuning, and low-rank adaptation.
The experimental setup for each approach is described below.

\textit{Direct prompting:} In this approach, we directly prompted a large scale LLM to predict the developer’s feedback. 
We evaluated two prompting strategies: zero‑shot and chain‑of‑thought~\cite{kojima2022large,wei2022chain}. 
Since this approach relies on the general pretrained knowledge of the underlying LLM, we experimented with one commercial model (GPT‑5.1)~\cite{singh2025openai} and six open‑source models ranging from 33B to 70B parameters.
These models were drawn from the LLaMA~\cite{grattafiori2024llama,roziere2023code}, Qwen~\cite{yang2025qwen3,cao2026qwen3}, and DeepSeek~\cite{guo2024deepseek,bi2024deepseek} families.
A temperature of 0 was used for all experiments to increase replicability.

\textit{Fine-tuning:} 
In this training-based approach, we employed the base and large variants of ModernBERT~\cite{warner-etal-2025-smarter}, a state-of-the-art encoder-only model, with parameter counts of 149M and 395M, respectively.
These small model sizes allowed us to train them by fully updating all parameters.
The use of a small encoder-only model aligns with prior work on classifying code review comments~\cite{10.1145/3611643.3616245}. 
ModernBERT, however, is a more advanced version, offering enhanced pretraining, improved architecture, and a significantly longer context window of 8,192 tokens compared to the 512 token limit of traditional BERT~\cite{devlin2019bert}. 
This extended context allows us to include both agentic code reviews and their corresponding code diffs in a single input.
Following established hyperparameters~\cite{warner-etal-2025-smarter,devlin2019bert}, we trained the models for three epochs with an effective batch size of 32, Bfloat16 precision, a learning rate of $2 e^{-5}$, and AdamW optimisation.

\textit{Low-rank adaptation:}
Unlike full fine-tuning, LoRA~\cite{lora} updates only low-rank adaptation matrices injected into specific layers, leaving the majority of model weights frozen.
This approach allowed us to train LLMs that are more parameterised, and presumably more capable than ModernBERT.
We considered lightweight variants from the same open-source families used in the direct prompting experiments.
Their parameter sizes range from 0.6B to 4B.
Following established hyperparameters~\cite{lora,dettmers2023qlora,lee2023platypus,ZhouLX0SMMEYYZG23}, we trained the models for three epochs with an effective batch size of 48, Bfloat16 precision, a learning rate of $2 e^{-4}$, a weighted decay of 0.01, and AdamW optimisation.
For LoRA, we used rank $r=16$, $\alpha=32$, 5\% dropout, and applied it to the query, key, value, output, gate, up, and down projection layers.

\underline{Experimental Setup:} For all experiments, we used the agentic code review as input, and tested both with and without the associated code diff.
For the test set, we used our manually annotated sample of 297 agentic code reviews. 
To train the models in fine‑tuning and LoRA experiments, we used the remaining 30,776 agentic code reviews that were labelled by GPT-5.1 in RQ1.
In total, we ran 40 experiments across the different approaches, model variants, families, and input contexts.
Finally, we measured the accuracy of predicting rejection using precision, recall, and F1 score.

\useunder{\uline}{\ul}{}

\begin{table}[!htbp]
\caption{Prediction Acc. of Rejected Agentic Code Reviews}
\label{tab:gating_results}
\resizebox{0.5\columnwidth}{!}{%
\huge
\sffamily 
\begin{tabular}{|lclrrr|}
\hline
\rowcolor[HTML]{EAEAEA} 
\textbf{Model} & \textbf{Context} & \textbf{Method} & \multicolumn{1}{c}{\cellcolor[HTML]{EAEAEA}\textbf{P}} & \multicolumn{1}{c}{\cellcolor[HTML]{EAEAEA}\textbf{R}} & \multicolumn{1}{c|}{\cellcolor[HTML]{EAEAEA}\textbf{F1}} \\ \hline
\rowcolor[HTML]{FFFFFF} 
\multicolumn{1}{|l|}{\cellcolor[HTML]{FFFFFF}} & \multicolumn{1}{c|}{\cellcolor[HTML]{EFEFEF}} & \multicolumn{1}{l|}{\cellcolor[HTML]{FFFFFF}Zero-Shot} & 0.54 & 0.05 & 0.08 \\
\rowcolor[HTML]{FFFFFF} 
\multicolumn{1}{|l|}{\multirow{-2}{*}{\cellcolor[HTML]{FFFFFF}GPT 5.1}} & \multicolumn{1}{c|}{\cellcolor[HTML]{EFEFEF}} & \multicolumn{1}{l|}{\cellcolor[HTML]{FFFFFF}Zero-Shot CoT} & 0.56 & 0.03 & 0.06 \\
\rowcolor[HTML]{EFEFEF} 
\multicolumn{1}{|l|}{\cellcolor[HTML]{EFEFEF}} & \multicolumn{1}{c|}{\cellcolor[HTML]{EFEFEF}} & \multicolumn{1}{l|}{\cellcolor[HTML]{EFEFEF}Zero-Shot} & 0.00 & 0.00 & 0.00 \\
\rowcolor[HTML]{EFEFEF} 
\multicolumn{1}{|l|}{\multirow{-2}{*}{\cellcolor[HTML]{EFEFEF}Llama-3.3-70B-Instruct}} & \multicolumn{1}{c|}{\cellcolor[HTML]{EFEFEF}} & \multicolumn{1}{l|}{\cellcolor[HTML]{EFEFEF}Zero-Shot CoT} & 0.00 & 0.00 & 0.00 \\
\rowcolor[HTML]{FFFFFF} 
\multicolumn{1}{|l|}{\cellcolor[HTML]{FFFFFF}} & \multicolumn{1}{c|}{\cellcolor[HTML]{EFEFEF}} & \multicolumn{1}{l|}{\cellcolor[HTML]{FFFFFF}Zero-Shot} & 0.53 & 0.23 & 0.32 \\
\rowcolor[HTML]{FFFFFF} 
\multicolumn{1}{|l|}{\multirow{-2}{*}{\cellcolor[HTML]{FFFFFF}CodeLlama-70b-Instruct-hf}} & \multicolumn{1}{c|}{\cellcolor[HTML]{EFEFEF}} & \multicolumn{1}{l|}{\cellcolor[HTML]{FFFFFF}Zero-Shot CoT} & 0.52 & 0.84 & 0.64 \\
\rowcolor[HTML]{EFEFEF} 
\multicolumn{1}{|l|}{\cellcolor[HTML]{EFEFEF}} & \multicolumn{1}{c|}{\cellcolor[HTML]{EFEFEF}} & \multicolumn{1}{l|}{\cellcolor[HTML]{EFEFEF}Zero-Shot} & 0.38 & 0.04 & 0.07 \\
\rowcolor[HTML]{EFEFEF} 
\multicolumn{1}{|l|}{\multirow{-2}{*}{\cellcolor[HTML]{EFEFEF}Qwen3-30B-A3B-Instruct-2507}} & \multicolumn{1}{c|}{\cellcolor[HTML]{EFEFEF}} & \multicolumn{1}{l|}{\cellcolor[HTML]{EFEFEF}Zero-Shot CoT} & 0.52 & 0.08 & 0.14 \\
\rowcolor[HTML]{FFFFFF} 
\multicolumn{1}{|l|}{\cellcolor[HTML]{FFFFFF}} & \multicolumn{1}{c|}{\cellcolor[HTML]{EFEFEF}} & \multicolumn{1}{l|}{\cellcolor[HTML]{FFFFFF}Zero-Shot} & 0.50 & 0.03 & 0.05 \\
\rowcolor[HTML]{FFFFFF} 
\multicolumn{1}{|l|}{\multirow{-2}{*}{\cellcolor[HTML]{FFFFFF}Qwen3-Coder-Next}} & \multicolumn{1}{c|}{\cellcolor[HTML]{EFEFEF}} & \multicolumn{1}{l|}{\cellcolor[HTML]{FFFFFF}Zero-Shot CoT} & 0.58 & 0.25 & 0.35 \\
\rowcolor[HTML]{EFEFEF} 
\multicolumn{1}{|l|}{\cellcolor[HTML]{EFEFEF}} & \multicolumn{1}{c|}{\cellcolor[HTML]{EFEFEF}} & \multicolumn{1}{l|}{\cellcolor[HTML]{EFEFEF}Zero-Shot} & 0.39 & 0.05 & 0.08 \\
\rowcolor[HTML]{EFEFEF} 
\multicolumn{1}{|l|}{\multirow{-2}{*}{\cellcolor[HTML]{EFEFEF}deepseek-llm-67b-chat}} & \multicolumn{1}{c|}{\cellcolor[HTML]{EFEFEF}} & \multicolumn{1}{l|}{\cellcolor[HTML]{EFEFEF}Zero-Shot CoT} & 0.56 & 0.26 & 0.35 \\
\rowcolor[HTML]{FFFFFF} 
\multicolumn{1}{|l|}{\cellcolor[HTML]{FFFFFF}} & \multicolumn{1}{c|}{\cellcolor[HTML]{EFEFEF}} & \multicolumn{1}{l|}{\cellcolor[HTML]{FFFFFF}Zero-Shot} & 0.45 & 0.32 & 0.38 \\
\rowcolor[HTML]{FFFFFF} 
\multicolumn{1}{|l|}{\multirow{-2}{*}{\cellcolor[HTML]{FFFFFF}deepseek-coder-33b-instruct}} & \multicolumn{1}{c|}{\multirow{-14}{*}{\cellcolor[HTML]{EFEFEF}CR}} & \multicolumn{1}{l|}{\cellcolor[HTML]{FFFFFF}Zero-Shot CoT} & 0.53 & 0.56 & 0.55 \\ \hline
\rowcolor[HTML]{EFEFEF} 
\multicolumn{1}{|l|}{\cellcolor[HTML]{EFEFEF}} & \multicolumn{1}{c|}{\cellcolor[HTML]{EFEFEF}} & \multicolumn{1}{l|}{\cellcolor[HTML]{EFEFEF}Zero-Shot} & 0.61 & 0.15 & 0.24 \\
\rowcolor[HTML]{EFEFEF} 
\multicolumn{1}{|l|}{\multirow{-2}{*}{\cellcolor[HTML]{EFEFEF}GPT 5.1}} & \multicolumn{1}{c|}{\cellcolor[HTML]{EFEFEF}} & \multicolumn{1}{l|}{\cellcolor[HTML]{EFEFEF}Zero-Shot CoT} & 0.63 & 0.11 & 0.19 \\
\rowcolor[HTML]{FFFFFF} 
\multicolumn{1}{|l|}{\cellcolor[HTML]{FFFFFF}} & \multicolumn{1}{c|}{\cellcolor[HTML]{EFEFEF}} & \multicolumn{1}{l|}{\cellcolor[HTML]{FFFFFF}Zero-Shot} & 0.00 & 0.00 & 0.00 \\
\rowcolor[HTML]{FFFFFF} 
\multicolumn{1}{|l|}{\multirow{-2}{*}{\cellcolor[HTML]{FFFFFF}Llama-3.3-70B-Instruct}} & \multicolumn{1}{c|}{\cellcolor[HTML]{EFEFEF}} & \multicolumn{1}{l|}{\cellcolor[HTML]{FFFFFF}Zero-Shot CoT} & 0.60 & 0.02 & 0.04 \\
\rowcolor[HTML]{EFEFEF} 
\multicolumn{1}{|l|}{\cellcolor[HTML]{EFEFEF}} & \multicolumn{1}{c|}{\cellcolor[HTML]{EFEFEF}} & \multicolumn{1}{l|}{\cellcolor[HTML]{EFEFEF}Zero-Shot} & 0.53 & 0.06 & 0.11 \\
\rowcolor[HTML]{EFEFEF} 
\multicolumn{1}{|l|}{\multirow{-2}{*}{\cellcolor[HTML]{EFEFEF}Qwen3-30B-A3B-Instruct-2507}} & \multicolumn{1}{c|}{\cellcolor[HTML]{EFEFEF}} & \multicolumn{1}{l|}{\cellcolor[HTML]{EFEFEF}Zero-Shot CoT} & \textbf{0.74} & 0.17 & 0.28 \\
\rowcolor[HTML]{FFFFFF} 
\multicolumn{1}{|l|}{\cellcolor[HTML]{FFFFFF}} & \multicolumn{1}{c|}{\cellcolor[HTML]{EFEFEF}} & \multicolumn{1}{l|}{\cellcolor[HTML]{FFFFFF}Zero-Shot} & 0.35 & 0.05 & 0.08 \\
\rowcolor[HTML]{FFFFFF} 
\multicolumn{1}{|l|}{\multirow{-2}{*}{\cellcolor[HTML]{FFFFFF}Qwen3-Coder-Next}} & \multicolumn{1}{c|}{\cellcolor[HTML]{EFEFEF}} & \multicolumn{1}{l|}{\cellcolor[HTML]{FFFFFF}Zero-Shot CoT} & 0.58 & 0.29 & 0.39 \\
\rowcolor[HTML]{EFEFEF} 
\multicolumn{1}{|l|}{\cellcolor[HTML]{EFEFEF}} & \multicolumn{1}{c|}{\cellcolor[HTML]{EFEFEF}} & \multicolumn{1}{l|}{\cellcolor[HTML]{EFEFEF}Zero-Shot} & 0.45 & 0.24 & 0.31 \\
\rowcolor[HTML]{EFEFEF} 
\multicolumn{1}{|l|}{\multirow{-2}{*}{\cellcolor[HTML]{EFEFEF}deepseek-coder-33b-instruct}} & \multicolumn{1}{c|}{\multirow{-10}{*}{\cellcolor[HTML]{EFEFEF}Diff \& CR}} & \multicolumn{1}{l|}{\cellcolor[HTML]{EFEFEF}Zero-Shot CoT} & 0.55 & 0.50 & 0.52 \\ \hline
\rowcolor[HTML]{FFFFFF} 
\multicolumn{1}{|l|}{\cellcolor[HTML]{FFFFFF}ModernBERT-base} & \multicolumn{1}{c|}{\cellcolor[HTML]{EFEFEF}} & \multicolumn{1}{c|}{\cellcolor[HTML]{EFEFEF}} & 0.52 & {\ul 0.96} & 0.68 \\
\rowcolor[HTML]{EFEFEF} 
\multicolumn{1}{|l|}{\cellcolor[HTML]{EFEFEF}ModernBERT-large} & \multicolumn{1}{c|}{\multirow{-2}{*}{\cellcolor[HTML]{EFEFEF}CR}} & \multicolumn{1}{c|}{\cellcolor[HTML]{EFEFEF}} & 0.53 & 0.94 & 0.68 \\ \cline{1-2} \cline{4-6} 
\rowcolor[HTML]{FFFFFF} 
\multicolumn{1}{|l|}{\cellcolor[HTML]{FFFFFF}ModernBERT-base} & \multicolumn{1}{c|}{\cellcolor[HTML]{EFEFEF}} & \multicolumn{1}{c|}{\cellcolor[HTML]{EFEFEF}} & 0.51 & 0.94 & 0.66 \\
\rowcolor[HTML]{EFEFEF} 
\multicolumn{1}{|l|}{\cellcolor[HTML]{EFEFEF}ModernBERT-large} & \multicolumn{1}{c|}{\multirow{-2}{*}{\cellcolor[HTML]{EFEFEF}Diff \& CR}} & \multicolumn{1}{c|}{\multirow{-4}{*}{\cellcolor[HTML]{EFEFEF}Fine-tuning}} & 0.52 & {\ul 0.96} & 0.68 \\ \hline
\rowcolor[HTML]{FFFFFF} 
\multicolumn{1}{|l|}{\cellcolor[HTML]{FFFFFF}Llama-3.2-1B} & \multicolumn{1}{c|}{\cellcolor[HTML]{EFEFEF}} & \multicolumn{1}{c|}{\cellcolor[HTML]{EFEFEF}} & 0.57 & 0.88 & 0.69 \\
\rowcolor[HTML]{EFEFEF} 
\multicolumn{1}{|l|}{\cellcolor[HTML]{EFEFEF}Llama-3.2-3B} & \multicolumn{1}{c|}{\cellcolor[HTML]{EFEFEF}} & \multicolumn{1}{c|}{\cellcolor[HTML]{EFEFEF}} & 0.62 & 0.79 & 0.70 \\
\rowcolor[HTML]{FFFFFF} 
\multicolumn{1}{|l|}{\cellcolor[HTML]{FFFFFF}Qwen3-0.6B-Base} & \multicolumn{1}{c|}{\cellcolor[HTML]{EFEFEF}} & \multicolumn{1}{c|}{\cellcolor[HTML]{EFEFEF}} & 0.58 & 0.92 & 0.71 \\
\rowcolor[HTML]{EFEFEF} 
\multicolumn{1}{|l|}{\cellcolor[HTML]{EFEFEF}Qwen3-1.7B-Base} & \multicolumn{1}{c|}{\cellcolor[HTML]{EFEFEF}} & \multicolumn{1}{c|}{\cellcolor[HTML]{EFEFEF}} & 0.63 & 0.78 & 0.70 \\
\rowcolor[HTML]{FFFFFF} 
\multicolumn{1}{|l|}{\cellcolor[HTML]{FFFFFF}Qwen3-4B-Base} & \multicolumn{1}{c|}{\cellcolor[HTML]{EFEFEF}} & \multicolumn{1}{c|}{\cellcolor[HTML]{EFEFEF}} & 0.66 & 0.83 & 0.73 \\
\rowcolor[HTML]{EFEFEF} 
\multicolumn{1}{|l|}{\cellcolor[HTML]{EFEFEF}deepseek-coder-1.3b-base} & \multicolumn{1}{c|}{\multirow{-6}{*}{\cellcolor[HTML]{EFEFEF}CR}} & \multicolumn{1}{c|}{\cellcolor[HTML]{EFEFEF}} & 0.59 & 0.78 & 0.67 \\ \cline{1-2} \cline{4-6} 
\rowcolor[HTML]{FFFFFF} 
\multicolumn{1}{|l|}{\cellcolor[HTML]{FFFFFF}Llama-3.2-1B} & \multicolumn{1}{l|}{\cellcolor[HTML]{EFEFEF}} & \multicolumn{1}{c|}{\cellcolor[HTML]{EFEFEF}} & 0.55 & \textbf{0.97} & 0.70 \\
\rowcolor[HTML]{EFEFEF} 
\multicolumn{1}{|l|}{\cellcolor[HTML]{EFEFEF}Llama-3.2-3B} & \multicolumn{1}{l|}{\cellcolor[HTML]{EFEFEF}} & \multicolumn{1}{c|}{\cellcolor[HTML]{EFEFEF}} & 0.65 & 0.86 & {\ul 0.74} \\
\rowcolor[HTML]{FFFFFF} 
\multicolumn{1}{|l|}{\cellcolor[HTML]{FFFFFF}Qwen3-0.6B-Base} & \multicolumn{1}{l|}{\cellcolor[HTML]{EFEFEF}} & \multicolumn{1}{c|}{\cellcolor[HTML]{EFEFEF}} & 0.64 & 0.79 & 0.70 \\
\rowcolor[HTML]{EFEFEF} 
\multicolumn{1}{|l|}{\cellcolor[HTML]{EFEFEF}Qwen3-1.7B-Base} & \multicolumn{1}{l|}{\cellcolor[HTML]{EFEFEF}} & \multicolumn{1}{c|}{\cellcolor[HTML]{EFEFEF}} & 0.65 & 0.80 & 0.72 \\
\rowcolor[HTML]{FFFFFF} 
\multicolumn{1}{|l|}{\cellcolor[HTML]{FFFFFF}Qwen3-4B-Base} & \multicolumn{1}{l|}{\cellcolor[HTML]{EFEFEF}} & \multicolumn{1}{c|}{\cellcolor[HTML]{EFEFEF}} & {\ul 0.68} & 0.87 & \textbf{0.76} \\
\rowcolor[HTML]{EFEFEF} 
\multicolumn{1}{|l|}{\cellcolor[HTML]{EFEFEF}deepseek-coder-1.3b-base} & \multicolumn{1}{l|}{\multirow{-6}{*}{\cellcolor[HTML]{EFEFEF}Diff \& CR}} & \multicolumn{1}{c|}{\multirow{-12}{*}{\cellcolor[HTML]{EFEFEF}LoRA}} & 0.66 & 0.80 & 0.73 \\ \hline
\end{tabular}
}
\end{table}

\textbf{Results.} 
Table \ref{tab:gating_results} shows the prediction results across different approaches.
For direct prompting, the top performer was CodeLlama 70b with zero-shot chain-of-thought, which achieved precision, recall, and F1 scores of 52\%, 84\%, and 64\%, respectively.
However, other models achieved far lower performances.
When using the agentic code review only, zero-shot achieved F1 scores between 0\% and 38\%, whilst zero-shot chain-of-thought achieved F1 scores between 0\% and 55\% (except the top performer).
Similarly, when using both agentic code review and the corresponding diff, zero-shot achieved F1 scores between 0\% and 31\%, whilst zero-shot chain-of-thought achieved F1 scores between 4\% and 52\%.
Even the highly-parameterised state-of-the-art model, GPT-5.1, achieved extremely low F1 scores ranging from 6\% to 24\%.
These results indicate that zero-shot predicting whether a code review will receive a rejection at inference time is unviable when the model is restricted to only the current review and its associated code diff. 

The results show that training‑based approaches, i.e., fine‑tuning and low‑rank adaptation, provided substantial improvements over direct prompting. Fine‑tuning ModernBERT yielded F1 scores of 66\%–68\%, outperforming the best direct‑prompting baseline by 2\%–4\%, despite using only 149M-395M parameters (0.2\%–0.6\% of CodeLlama‑70B). 
Similarly, applying low‑rank adaptation to moderately larger models achieved F1 scores of 67\%–76\%, surpassing direct prompting by 3\%–12\%. These models remained compact, ranging from 0.6B to 4B parameters (0.9\%–5.7\% of CodeLlama‑70B).
The strongest overall performance came from Qwen3‑4B using both the code review and its corresponding code diff as input, achieving 68\% precision, 87\% recall, and a 76\% F1 score. Llama3.2‑3B with the same input setting achieved the second‑best results, with 65\% precision, 86\% recall, and a 74\% F1 score. 
The gains over direct prompting are driven primarily by large improvements in precision (16\% and 13\%), with recall increasing more modestly by 3\% and 2\%.
These results suggest that prior developer feedback encode learnable rejection patterns even within limited context i.e., the review and associated code diff, enabling lightweight LLMs to address the precision bottleneck of direct-prompting approaches, without requiring the high overhead of retrieving and reasoning over additional context.

\summary{
Direct prompting with large LLMs only achieves F1 scores up to 64\% in developer rejection prediction, while training-based approaches, particularly low-rank adaptation on small LLMs, can achieve up to 76\%, suggesting that meaningful patterns can be extracted from prior developer feedback, even under limited context. 
This represents a viable lightweight approach to inference time quality filtering of agentic code reviews.
}

\section{Discussion \& Recommendations}
We now further discuss opportunity gaps and shortcomings, as well as offer recommendations.

\textbf{Opportunity Gaps for CodeRabbit's Agentic Code Reviews.}
Findings from RQ2 show long-tail evolvability-related categories ($\leq$0.1\%) e.g., naming convention, visual representation, and testing, with higher acceptance rates (60.0\%–87.5\%).
Whether this reflects the underlying distribution of review opportunities or a tendency of the agents to underproduce such suggestions remains an open question.
While naming convention and visual representation may be considered relatively trivial~\cite{turzo2024makes}, valid instances can still improve the long-term maintainability and consistency of the codebase.
For example, reviews targeting naming convention suggest ways to improve understandability of the code: \textit{``The test method testGetApplication\_success should be renamed to testGetOrders\_success to reflect the updated functionality.''}, while reviews targeting visual representation suggest ways to improve code readability: \textit{``Normalize link text and anchors for prerequisites \& style guide''}.
These reminders can help prevent the gradual accumulation of technical debt.
Additionally, CodeRabbit effectively identified missing test coverage and provided actionable suggestions, such as \textit{``Consider adding unit tests for merged case scenarios''}.
Although not directly improving the code in its current state, these suggestions function as important reminders to revisit test coverage and improve robustness against edge cases.

Rejected evolvability reviews were more likely to be technically valid but dismissed only due to misalignments with coding practices (31.6\% vs. 25.1\%).
Notably, most rejected suggestions related to solution approach and organisation of code are attributed to this reason, accounting for 34.7\% and 30.7\% of each case, respectively.
For example, a valid review offering an improved implementation: \textit{``Here's a more concise alternative using split()''} may be dismissed due to personal preferences for maintaining understandability: \textit{``A less concise version which showcases how to handle the language primitives might be more appropriate here''}.
Similarly, valid reviews offering better code structure, such as: \textit{``The conditional logic for fileManager and associating\_id could be simplified for better maintainability.''}, could be unwarranted in the current phase of development: \textit{``Shouldn't be an issue for now ?''}.
These findings suggest an opportunity gap where, despite being technically more competent in these areas, the agentic code reviews require stronger personalisation to better align evolvability suggestions with individual developer practices, preferences, and development contexts.

\textbf{Shortcomings of CodeRabbit's Agentic Code Reviews.}
Our RQ1 results show that 58.0\% of agentic code reviews were rejected as invalid, most commonly due to false-positive issues (Table~\ref{tab:rejection_reasons}). Our coding exercise also surfaced several recurring patterns behind these failures.
Among the 7,570	reviews classified as false positives, 43.0\% could be attributed to the agent’s limited understanding of the system’s overall design and functional behaviour. 
For example, CodeRabbit produced the resource-related recommendation: \textit{“Security: don’t set state to slack\_user\_id”}, which the developer rejected, noting the system design: \textit{“But I will need the slack\_user\_id parameter later in the callback”}.
We further observed that in 43.4\% of false-positive cases, the agent fails to understand the specific section of code that is directly under review.
For example, an agentic code review flagged a functional defect: 
\textit{``The helper function uses TIMEOUT\_2MIN and TimeoutSampler, but neither is imported, which will cause runtime errors''}.
The developer clarified that this was a misunderstanding of the code implementation: 
\textit{``The function that used them has been removed, so these imports are no longer needed.''}
In a smaller number of cases (10.3\%), the agent lacks visibility into inter‑file code implementations, resulting in misidentified issues.
For example, an agentic code review flagged a functional defect: \textit{``no matching handlers or validators were found ... Please ensure these new endpoint types are wired end-to-end''}, which resulted in the developer informing: \textit{``The isValidWSAPI() method in PublisherCommonUtils handles validation''}, which invalidates the suggestion.
The remaining 3.3\% of false positive issues were due to lack of oversight of the project’s software development lifecycle.

Similarly, we observed that the 822 redundant suggestions and 1,763 out of scope issues stem from shortcomings in the agent's understanding of the system's design and functionality (9.3\%, 35.1\%), incorrect program analysis of the current file (17.8\%, 7.3\%), restricted inter-file visibility (38.1\%, 5.2\%), as well as lack of consideration for the development lifecycle (34.8\%, 52.4\%), including concurrent PRs and open issue tickets that were previously raised.
In RQ2, the high rate of invalid suggestions targeting functional defect, interface, resource, timing, and alternate output (36\%-43\%) further confirms that non‑helpful reviews frequently stem from limitations in the agent’s understanding of high‑level system architecture, runtime behavior, repository‑wide context, as well as complex code structures.
These findings highlight a limitation in the agent’s ability to comprehend the system in its entirety, ultimately resulting in invalid code reviews.

\textbf{Human vs. CodeRabbit's Agentic Code Reviews.}
The stronger emphasis on identifying functional defects (43.3\%) contrasts notably with traditional human-conducted code reviews, where such comments are rare. 
Notably, there is a well-documented gap between developers’ expectations that code reviews will uncover defects and the actual proportion identified by human reviewers~\cite{7202946,hasan2021using,6606617}.
A widely cited statistic from Microsoft suggests that only 15\% of reviews identify a potential defect~\cite{7202946}, whereas evolvability-related reviews constitute the majority.
When valid, these type of reviews are commonly considered the most useful~\cite{hasan2021using,turzo2024makes,7180075}, but usually require experienced human reviewers with an in-depth understanding of the code~\cite{6606617,10.1145/2884781.2884852,7332457,10.1145/2884781.2884840}.
Against this backdrop, the 36.2\% of functional defects raised and accepted by developers, as well as the 7.3\% that prompted discussion, serve as valuable complements to human reviews.
As such, a key trade-off is between the value contributed by these additionally identified defects and the productivity cost to developers who must waste time and effort inspecting the 35.8\% of false positive issues.

\textbf{Recommendations.} We now offer recommendations to various stakeholders in the software engineering community. 
\begin{enumerate}
    \item \emph{For tool builders in agentic code review}, our findings first highlight the need to reduce fundamental program understanding errors with the specific lines of code under inspection. 
    Beyond this, the agentic code review tool requires a deeper understanding of how system components are integrated across the repository, as well as enabling inspection of system behaviour at runtime.
    Additionally, the agentic code review tool should better leverage historical software development artefacts to align with the developers' coding practices.
    
    \item \emph{For developers using agentic code review tools}, our findings highlight the strengths and weaknesses of agentic code reviews in its current form, helping to better align expectations of these tools.
    If there is a strong preference for reducing agentic reviews that are ultimately rejected, a predictive approach to gate non-helpful reviews could help.
    Our RQ3 findings indicate that feedback patterns to agentic reviews are learnable and can be used to early predict whether an agentic code review will be rejected by developers.

    \item \emph{For researchers}, our findings suggest that developer feedback provide meaningful insights of code review quality. Historically, research on automated code review has relied primarily on offline evaluation using reference-based metrics (e.g., exact match, BLEU~\cite{10378848,10.1145/3762183,li2022automating}), largely because automated review systems had limited real-world adoption. 
    As agentic code review tools like CodeRabbit become widely deployed and generate large volumes of developer interaction data, future work could leverage developer feedback as an additional evaluation signal, complementing traditional offline metrics and providing a measure that better reflects practical usefulness.
\end{enumerate}

\section{Threats to the Validity}
We discuss the internal and external threats to validity in our study.

\textbf{Internal Validity.}
The manual categorisation of developer feedback and review types (RQ1 and RQ2) involves a degree of subjectivity.
To mitigate this threat for RQ1, the open coding process was conducted iteratively by two authors through multiple rounds of discussion and refinement.
To mitigate this threat for RQ2, the manual annotation was conducted independently by two authors, achieving substantial inter-rater agreement (Cohen's $k=0.85$).
Any discrepancies were resolved with a third author.
All final annotations were verified by the second author.
For the remaining data, GPT-5.1 was employed as a state-of-the-art LLM annotator, which may introduce label noise.
To assess this risk, we compared its annotations against our manually derived labels for both RQ1 and RQ2, observing substantial agreement (Cohen's $k=0.74$ and $0.78$, respectively).
The strong performance of the supervised approaches in RQ3 further suggests that the remaining annotations are sufficiently reliable for our analysis.
While prior research on automated code review tools deployed in practice has often used subsequent code changes as a proxy for developer acceptance~\cite{tantithamthavorn2026rovodev,11495549}, our study instead focuses on the distribution of the developers' explicit feedback. 
Thus, the content of developers' responses alone provides a sufficient signal for our use case. 
Moreover, this approach captures scenarios where developers indicate that they intend to address the issue in a subsequent PR or by creating a new issue. 
Such cases would be misclassified as false negatives when relying solely on signals such as subsequent commits.

\textbf{External Validity.}
Our study focuses exclusively on CodeRabbit. 
While it is a highly popular agentic code review tool, our findings may not completely generalise to other agentic review assistants that utilise different underlying models, prompts, or context-enrichment architectures.
Nevertheless, because CodeRabbit embodies common design patterns, our results capture behaviours that are likely to be shared across a broader class of such tools.
Our dataset focuses on open-source GitHub projects with permissive licenses across the ten most popular programming languages. 
Consequently, our findings may not generalise to industrial, closed-source environments or to niche, legacy languages.
To ensure broad applicability, we explicitly selected languages that account for the vast majority (77\%) of real-world usage.
Furthermore, we applied stratified sampling across both languages and repositories to preserve representative distributional characteristics, and filtered out projects with minimal agent interaction to ensure our data reflects sustained, practical usage in routine software engineering rather than experimental trials.

\section{Conclusion}
This paper is the first to present an in-depth empirical study on CodeRabbit's agentic code reviews, grounded in developer feedback and the underlying reasons behind them. 
Our findings indicate that agentic reviews elicit mixed feedback, with a 36.4\% acceptance rate, a 56.3\% rejection rate, and 7.3\% of cases triggering a discussion.
On a positive note, we find that the acceptance rate has been steadily increasing since release, reflecting practical improvements of the system.
Rejections were mainly driven by misalignments with coding practices and invalid suggestions, with an increasing representation of the former among code reviews with negative feedback. This suggests that CodeRabbit is become more technically proficient, but still struggles to conform to developers' practices.
Our findings show that CodeRabbit has a growing focus on functional concerns, specifically identifying functional defects, but still have higher rejection rates due to invalidity of the suggestions.
Through a systematic evaluation of LLM-based approaches, we find that lightweight LLMs are able to learn patterns aligning agentic code reviews to developer feedback, allowing for early prediction of rejection.
For developers who want to reduce the amount of code reviews that would be ultimately rejected, this method can be used as a gating mechanism to reduce the burden of inspecting reviews of lower quality.

\section{Data Availability}
Our replication package, including datasets, manual and automatic labelling results, model training, inference, and evaluations scripts is available at: https://doi.org/10.5281/zenodo.21095394

\bibliographystyle{ACM-Reference-Format}
\bibliography{references}

\end{document}